\documentclass[fleqn,11pt]{wlscirep}

\usepackage[utf8]{inputenc}
\usepackage[T1]{fontenc}
\usepackage[english]{babel}

\usepackage{amsmath}
\usepackage{amssymb}
\usepackage{siunitx}
\usepackage{float}
\usepackage{graphicx}
\usepackage{tikz}
\usepackage{epsfig}
\usepackage{booktabs}
\usepackage{longtable}
\usepackage{colortbl}
\usepackage{multirow}
\usepackage{makecell}
\usepackage{placeins}
\usepackage{xcolor}
\usepackage{caption}
\usepackage{hyperref}
\usetikzlibrary{decorations.pathmorphing}

\definecolor{MerminBlue}{HTML}{4A90D9}
\definecolor{SvetGold}{HTML}{E8A838}
\definecolor{lightred}{RGB}{255, 200, 200}
\definecolor{lightblue}{RGB}{200, 200, 255}
\definecolor{cyan}{RGB}{0, 255, 255}
\definecolor{royalblue}{RGB}{65, 105, 225}
\definecolor{lightblue3}{RGB}{162, 181, 205}
\definecolor{lightgray}{gray}{0.9}

\usepackage[ruled,vlined]{algorithm2e}
\usepackage{verbatim}
\usepackage{lipsum}
\usepackage{enumitem}
\usepackage{orcidlink}
\usepackage{array}
\usepackage{cuted}
\newcolumntype{P}[1]{>{\centering\arraybackslash}p{#1}}

\usepackage{subcaption}
\usepackage{array}
\usepackage{makecell} 
\usepackage{array}

\newcolumntype{M}[1]{>{\centering\arraybackslash}m{#1}}
\usepackage{comment}
\usepackage{braket}

\newcolumntype{C}[1]{>{\centering\arraybackslash}p{#1}}

\usepackage{hyperref}
\usepackage{nameref} %
\makeatletter
\renewcommand\@makefnmark{%
    \hbox{\textsuperscript{\textcolor{red}{\@thefnmark}}}} 
\makeatother

\usepackage{todonotes}
\usepackage{soul}

\renewcommand{\textcolor}[2]{#2}

\renewcommand{\part}[1]{\left(#1\right)}

\usepackage{amsthm}

\usepackage{titlesec}

\renewcommand{\thesection}{\Roman{section}}
\renewcommand{\thesubsection}{\Alph{subsection}}
\renewcommand{\thesubsubsection}{\arabic{subsubsection}}



\titleformat{\section}
  {\color{color1}\large\sffamily\bfseries\centering}
  {\thesection.}
  {0.5em}
  {\MakeUppercase{#1}}

\titleformat{name=\section,numberless}
  {\color{color1}\large\sffamily\bfseries\centering}
  {}
  {0em}
  {\MakeUppercase{#1}}

\titleformat{\subsection}
  {\sffamily\bfseries\centering}
  {\thesubsection.}
  {0.5em}
  {\MakeUppercase{#1}}

\titleformat{\subsubsection}
  {\sffamily\small\bfseries\itshape}
  {\thesubsubsection.}
  {0.5em}
  {#1}

\begin{abstract}
        Benchmarking quantum processors is a central task in every roadmap aimed at quantum advantage. Here, we discuss a benchmarking protocol based on the violation of the Svetlichny inequality, which is designed to \textcolor{red}{simulate} genuine multipartite nonlocality. We compare it with Mermin's inequality, and present the results of applying this framework to two different quantum processing units —the superconducting system Garnet (from IQM) and the ion-trap system Marmot (from AQT), \textcolor{red}{ testing different families of quantum circuits for three to five qubits under identical experimental conditions. We also run simultaneous triplet tests across the full register, including a comparison between GHZ and linear cluster states, to probe how Svetlichny-inequality violations depend on qubit location and calibration}. Our results demonstrate that tests based on genuine multipartite nonlocality provide a powerful tool to characterize the collective performance of quantum devices and to guide their calibration and optimization.
\end{abstract}

\begin{document}

\twocolumn

\title{Nonlocality as a  Benchmarking Tool for Quantum Processors}

\author[1]{Andrés Camilo Granda Arango\orcidlink{0000-0002-5255-2507}}

\author[2,3]{Carla Sophie Rieger\orcidlink{0000-0002-9263-8604}}

\author[1]{Carlo Cuccu\orcidlink{0000-0002-4519-1854}}

\author[1]{Giuseppe Sergioli\orcidlink{0000-0002-3650-5858}}

\author[4]{Luigi Iapichino\orcidlink{0000-0003-3938-8973}}

\author[4]{Tobias Valentin Bauer\orcidlink{0009-0004-8511-0563}}

\author[1,5]{Roberto Giuntini\orcidlink{0000-0003-2078-0297}}

\author[1,6,*]{Federico Hernán Holik\orcidlink{0000-0002-6776-5281}}

\affil[1]
{\small
\href{https://ror.org/003109y17}{Università degli Studi di Cagliari}, Via Is Mirrionis 1, Cagliari, 09123, Italy}

\affil[2]
{\small
Department of Engineering and Design,
\href{https://ror.org/02kkvpp62}{Technische Universität München}, Lise-Meitner-Str. 9, 85521 Ottobrunn, Germany}

\affil[3]{\small
\href{https://ror.org/01ggx4157}
{European Organization for Nuclear Research (CERN)}, Espl. des Particules 1, 1211 Meyrin, Switzerland}

\affil[4]
{\small
\href{https://ror.org/05558nw16}{Leibniz Supercomputing Centre
of the Bavarian Academy of Sciences and Humanities (LRZ)}, Boltzmannstr. 1, 85748 Garching b. M\"unchen, Germany}

\affil[5]{\small
Institute for Advanced Study (IAS), \href{https://ror.org/02kkvpp62}{Technische Universit\"at M\"unchen}, Lichtenbergstraße 2a, 85748. M\"unchen, Germany}

\affil[6]{\small
Instituto de F\'isica (IFLP-CCT-CONICET), \href{https://ror.org/01tjs6929}{Universidad Nacional de La Plata}, C.C. 727, La Plata, 1900, Argentina}

\affil[*]{\small Contact author: holik@fisica.unlp.edu.ar}


\maketitle
\section{INTRODUCTION}

Quantum computers hold the promise to perform computational tasks more efficiently than their classical counterparts for specific problem 
classes~\cite{Deutsch1992,Shor1997,Simon1997,Riste2017}. In particular, they are expected to simulate many-body quantum systems, a problem that is believed to be exponentially hard for classical  algorithms~\cite{Feynman1982,Lloyd1996,Bernien2017,Cubitt2018}. As early examples of quantum processors are already available on the market, the simulation of quantum systems has yielded very promising proof-of-concept results~\cite{Bloch2012,Bernien2017, Abanin2025}. \textcolor{blue}{This line of research is particularly relevant because ground states of physically interesting many-body Hamiltonians can exhibit correlations that violate multipartite Bell
inequalities~\cite{Tura2014}}.

Before addressing the simulation of complex Hamiltonians, one may ask a more elementary question: up to which point are existing quantum processors capable of generating correlations \textcolor{blue}{simulating} multipartite Bell inequalities?~\cite{Mermin_1990,Svetlichny_PRD,Seevinck_Svetlichny_PRL}

\textcolor{blue}{While several studies have addressed this problem using Bell inequalities that do not exclude hybrid local realism (HLR)~\cite{Cabello-Non-locality-benchmarking,Lanyon2014}, others have proposed inequalities capable of certifying genuine multipartite nonlocality (GMNL) as alternative benchmarking tools~\cite{Baumer2021,Wang_2025,Granda_2025}. In this work, we focus on the Svetlichny inequalities, which are designed to certify GMNL by excluding all hybrid local-realistic models~\cite{Seevinck_Svetlichny_PRL}. This makes them particularly attractive as a benchmarking tool for quantum processors because \textcolor{red}{simulating} their violation involves the computation of correlations that go beyond those detected by standard multipartite Bell inequalities.}

\textcolor{blue}{Multiple protocols exist for characterizing quantum processor performance, including randomized benchmarking~\cite{Magesan2011}, quantum volume~\cite{cross2019}, cross-entropy benchmarking (XEB)~\cite{Arute_2019_Sycamore}, cycle benchmarking~\cite{erhard2019characterizing}, and mirror circuits~\cite{proctor2022scalable}. Building upon the framework established in~\cite{Granda_2025}, we further develop a protocol that evaluates a quantum processing unit (QPU) based on its ability to simulate genuine multipartite nonlocality. Starting from an initial stage of quantum random circuits, our benchmark quantifies the processor's capability by the degree to which its output correlations violate specific inequalities originally tailored to detect genuine multipartite nonlocality. Due to the fact that all qubits in a quantum computer are colocated, ensuring space-like separation between measurements is impossible; consequently, the resulting correlations suffer from loopholes that preclude a strict assertion of genuine multipartite nonlocality (see discussion in Refs.~\cite{Cabello-Non-locality-benchmarking,Granda_2025}). Nonetheless, as we demonstrate below, the task of simulating a nonlocality test on a quantum computer remains highly informative of its overall performance. Furthermore, as discussed in Ref.~\cite{Granda_2025} Section VI B, the correlations generated via this approach can be formally reinterpreted as a form of contextuality, thereby serving as a direct metric of the device's non-classical resources. It is important to recall here that contextuality has been targeted as a condition for quantum advantage in relevant quantum computing frameworks \cite{Howard2014}}.

\textcolor{red}{The theoretical foundations of the nonlocality test investigated here were first introduced in Ref.~\cite{Granda_2025}, which included preliminary experimental evaluations on the IQM Garnet and IonQ Aria-1 processors. To establish this technique as a viable, practical diagnostic tool, it is essential to provide comprehensive evidence of its real-world utility. In this work, we deploy the methodology in a systematic manner, significantly expanding both the numerical and experimental data necessary to demonstrate its power as a reliable benchmark. Furthermore, fully realizing the practical value of this technique requires a deep understanding of its comparative advantages; to this end, 
we provide rigorous evidence that the Svetlichny test is more stringent than the Mermin test: its simulated violations degrade more rapidly under depolarizing noise, and we identify explicit experimental regimes where the Mermin inequality is violated while the Svetlichny inequality remains satisfied—insights that are indispensable for demonstrating the tool's true capabilities in assessing and optimizing real-world quantum hardware.}

We test the protocol on two architecturally distinct processors operated at the Leibniz Supercomputing Centre (LRZ) in Garching, Germany: the sixteen-qubit\footnote{At the time the experiments reported in this work were performed, only 12 of the 16 qubits of the Marmot device were available for user access.} ion-trap system manufactured by Alpine Quantum Technologies (AQT), hereafter referred to as Marmot~\cite{AQT_Marmot}, and the twenty-qubit superconducting system developed by IQM, referred to as Garnet~\cite{Garnet_benchmark}. Both processors are evaluated under identical logical circuits and a common experimental protocol, \textcolor{red}{testing GHZ states and two families of random circuits: one built with gates from the Clifford group (\texttt{H}, \texttt{S}, and \texttt{CNOT}), which is finite and not universal; in Ref.~\cite{Granda_2025} it was verified that random sampling reproduces the distribution of violations obtained from the complete enumeration. The second family uses the same Clifford gates but adds the \texttt{T} gate, which makes the gate set universal and allows the random circuits, at sufficient depth, to approximate the Haar measure on the unitary group. The three families probe how the structure of the prepared state interacts with hardware noise on each platform. We also investigate the spatial dependence of simulated Svetlichny-inequality violations within each processor, running simultaneous triplet tests under full-register load on Marmot and comparing different qubit subsets with native gates at different calibration times on Garnet. Quantitative evaluation metrics are defined to allow direct comparison between the two architectures.} All experiments were carried out between December 2024 and June 2025.

The paper is organized as follows. Section~\ref{s:Framework} introduces the Svetlichny and Mermin inequalities.
\textcolor{red}{There, we show numerical evidence that violation of the Svetlichny inequality is more stringent than that of Mermin's. We also compare the nonlocality test with the performance of a particular quantum algorithm}. Section~\ref{s:Experiments} introduces the experimental methodology and reports the results of the nonlocality tests performed on the Marmot and Garnet processors, showing how the proposed approach can be used to assess and compare the performance of different architectures. 
In Section~\ref{s:Conclusions}, after making some remarks about the interpretation of the experimental results, we draw our conclusions. Finally, in Appendix~\ref{appendix:shots}, we examine the statistical stability of the measured Svetlichny violations
as a function of the number of measurement shots. Appendix~\ref{appendix:metrics} summarizes the quantitative evaluation measures used throughout the work to compare experimental data, simulations, and theoretical predictions.

\section{Nonlocality Framework}
\label{s:Framework}

\subsection{Svetlichny and Mermin Inequalities}

We first consider the bipartite Bell scenario described by the Clauser--Horne--Shimony--Holt (CHSH) inequality~\cite{CHSH_inequalities}. In this case, two spatially separated systems are measured using dichotomic observables, and the resulting correlations are compared with the predictions of local hidden-variable models. A violation of the CHSH inequality indicates bipartite quantum nonlocality, excluding any local realistic description of the observed correlations.

When extending this analysis to systems composed of more than two parties, the question arises as to whether multipartite correlations can still be explained within a local hidden-variable framework.
This problem was addressed by Mermin~\cite{Mermin_1990} (see also~\cite{Alsina_2016} and~\cite{Werner_2001}). By requiring the condition of the local hidden variable (LHV) model, it is possible to derive a set of inequalities based on the so-called
Mermin polynomials $M_{N}$ (see Section~II-C of ~\cite{Alsina_2016}).
For each party $j$, two dichotomic observables, denoted $A_j$ and
$A'_j$, are considered. The polynomials are recursively defined as: 

\begin{equation}
    M_{2}=\frac{1}{2}\left ( A_{1}A_{2}+ A_{1}'A_{2}+ A_{1} A_{2}' -A_{1}'A_{2}' \right ),
\end{equation}

and

\begin{align}
    M_{N}=\frac{1}{2}M_{N-1}\left ( A_{N}+A_{N}' \right )+\frac{1}{2}{M}_{N-1}'\left ( 
A_{N}- A_{N}'\right ),
\end{align}

\noindent where $M'_{N-1}$ is obtained from $M_{N-1}$ by exchanging $A_j$
and $A'_j$ for all parties. All local hidden variable theories obey the following inequality~\cite{Alsina_2016}:

\begin{equation}\label{e:Mermin}
 |\langle M_{N} \rangle| \leq 1.
\end{equation}

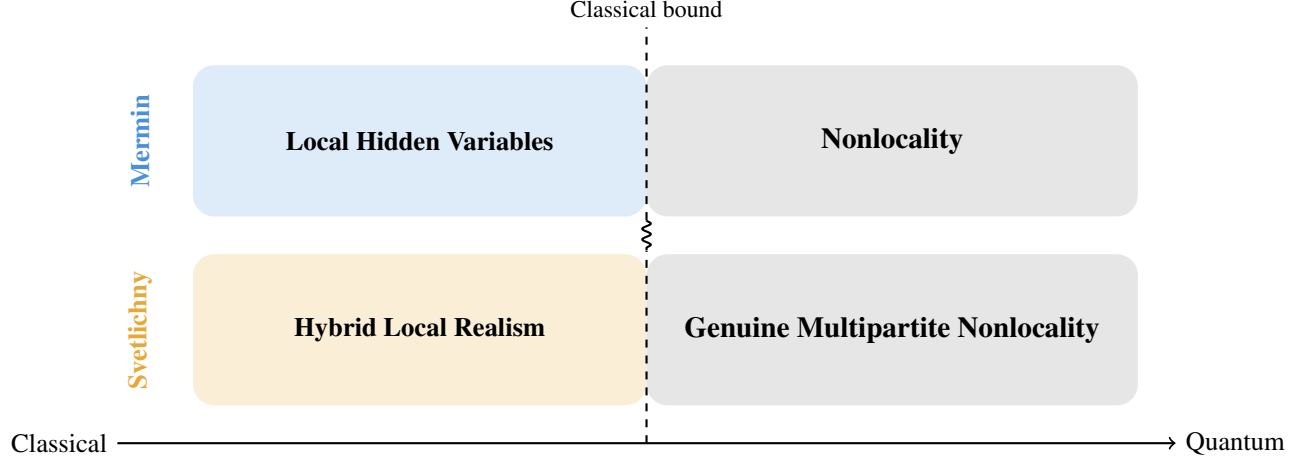
\begin{figure*}[hbt!]
    \centering
    \begin{tikzpicture}[font=\small]

    \fill[MerminBlue!18, rounded corners=8pt] (-6,2.5) rectangle (0,4.5);
    \node at (-3,3.5) {\textbf{Local Hidden Variables}};

    \fill[gray!20, rounded corners=8pt] (0,2.5) rectangle (6.5,4.5);
    \node[font=\bfseries] at (3.25,3.5) {Nonlocality};

    \node[
        anchor=center,
        rotate=90,
        font=\small\bfseries,
        text=MerminBlue
    ] at (-6.7,3.5) {Mermin};

    \fill[SvetGold!20, rounded corners=8pt] (-6,0) rectangle (0,2);
    \node at (-3,1) {\textbf{Hybrid Local Realism}};

    \fill[gray!20, rounded corners=8pt] (0,0) rectangle (6.5,2);
    \node[font=\bfseries] at (3.25,1) {Genuine Multipartite Nonlocality};

    \node[
        anchor=center,
        rotate=90,
        font=\small\bfseries,
        text=SvetGold
    ] at (-6.7,1) {Svetlichny};

    \draw[dashed, thick] (0,-0.5) -- (0,5);
    \node[font=\footnotesize, anchor=south] at (0,5) {Classical bound};

    \fill[white] (-0.15,2.05) rectangle (0.15,2.45);
    \draw[thick, decorate,
          decoration={snake, amplitude=1.6pt, segment length=4pt}]
          (0,2.45) -- (0,2.05);

    \draw[->, thick] (-7,-0.5) -- (7,-0.5);
    \node[anchor=east] at (-7,-0.5) {Classical};
    \node[anchor=west] at (7,-0.5) {Quantum};

    \end{tikzpicture}

    \caption{Regions defined by the classical bound of each
    inequality. For Mermin, exceeding the bound indicates
    incompatibility with fully local hidden-variable models.
    For Svetlichny, it indicates incompatibility with all hybrid
    local-realistic decompositions, consistent with genuine
    multipartite nonlocality.}
    \label{fig:quantumness-region}
\end{figure*}

\noindent If the correlations associated with a given quantum state are such that the above inequality is violated, then there exists no local hidden variable model for them.

A violation of Mermin's inequality thus excludes fully local
hidden-variable models. However, for multipartite systems weaker
forms of nonlocality remain possible. As an example, consider the three-particle state:

\begin{equation}\label{e:hybrid_local}
    |\psi\rangle = \frac{1}{\sqrt{2}}(|01\rangle-|10\rangle)|0\rangle
\end{equation}

\noindent which is a product of a two-qubit singlet state for the first two particles, and a pure state for the third particle. As such, the correlations of the first two particles yield maximal violation of the CHSH inequality, but the first two particles and the third one are completely disentangled. Such states illustrate that multipartite systems may contain nonlocal correlations within a subset of parties while still admitting an HLR description at the global level. Svetlichny inequalities were developed to discard such hybrid models.

\textcolor{blue}{An HLR model assumes that the observed correlations can be decomposed as a convex mixture over bipartitions of the $N$ parties: each term admits a local-realistic description across one bipartition, while nonlocal correlations may be present within either group. This coexistence of local and nonlocal elements across different subsets is what makes the model hybrid. A violation of the Svetlichny inequality rules out all such decompositions simultaneously, implying that the nonlocal correlations are genuinely multipartite rather than confined to any subset of parties.}

For the case of an arbitrary number $N$ of qubit systems, the Svetlichny inequality is built using the \textcolor{blue}{operator}~\cite{Seevinck_Svetlichny_PRL}:

\begin{equation}\label{e:Svelichny_N_Qubits_operator}
    S_{N}^{\pm}=\sum_{I}\nu _{t\left ( I \right )}^{\pm}A^{i_{1}}_{1}\cdots A^{i_{N}}_{N},
\end{equation}

\noindent where
$\nu_{t}^{\pm} = (-1)^{\frac{t(t \pm 1)}{2}}$, with $I = (i_{1}, i_{2}, \dots, i_{N})$, and each $i_{j}$ indicating whether a prime appears or not in the corresponding variable. The number $t\left ( I \right )$ indicates the number of times primes appear in $I$, and $\nu_{t}^{\pm}$ is a sequence of signs. Under the assumption of HLR~\cite{Svetlichny_PRD,Seevinck_Svetlichny_PRL}, two types of inequalities are derived, namely

\begin{equation}\label{e:SVL_N_Qubits}
    |S_{N}^{\pm}|\leq 2^{N-1}.
\end{equation}

\noindent If the correlations associated with a given quantum state are such that inequality~\ref{e:SVL_N_Qubits} is violated, then they cannot be modeled using HLR models. In other words, one can say that the correlations 
display \textit{genuine multipartite nonlocality}. As an example, the correlations associated with the GHZ state yield a violation of Eq.~\eqref{e:SVL_N_Qubits}.


\textcolor{red}{These two levels of nonlocality are summarized in 
Fig.~\ref{fig:quantumness-region}. For the analysis that 
follows, we refer to the region beyond the classical bound 
of each inequality as the \emph{nonclassical region}.}

\textcolor{blue}{We note that the Svetlichny inequality provides a sufficient condition for GMNL; the absence of a violation does not exclude genuine multipartite nonlocal correlations that might be detected by other inequalities.}

\textcolor{blue}{The violation of inequalities such as Mermin's does not guarantee the presence of genuine multipartite nonlocal correlations. The Svetlichny inequalities, by contrast, are specifically designed to discard all HLR models~\cite{Seevinck_Svetlichny_PRL}. Since GMNL
is by definition a property of the correlations of the system as a
whole, the Svetlichny test is a natural candidate for benchmarking
the collective performance of a quantum processor; we therefore
adopt it as the basis for our protocol, described below.} 

\textcolor{blue}{As mentioned in the Introduction, the correlations associated to the Svetlichny inequality considered in this work are evaluated on quantum processors where qubits are confined in a reduced space. No space-like separation is granted for measurements involving different qubits, and therefore, avoiding the locality loophole on quantum processors remains a challenging issue~\cite{Aspect1982,Hensen_2015,Significant-Loophole-Free-2015,Strong-Loophole-Free-2015,Rosenfeld2017}. For this reason, the protocol based on the Svetlichny inequality cannot be used to certify nonlocality in a loophole-free sense. As discussed in  Ref.\cite{Granda_2025} (Sec.VI.B), the violation of nonlocality inequalities on a QPU can be better understood as a contextuality phenomenon, since nonlocality inequalities are a particular case of noncontextuality inequalities\cite{Howard2014}. Following Ref.\cite{Granda_2025}, we speak of \textit{simulating nonlocality} to describe what the proposed test does. This means to reproduce the correlations associated to all the terms involved in the Svetlichny and Mermin's inequalities on a quantum processor. As we show in this work, it turns out that simulating those correlations turns out to be a hard task, that can be use to assess the collective behavior of different subsets of qubits. The test is so stringent, that it allows to detect differences in performance between quantum processors even when a small number of qubits is involved.} 

\textcolor{blue}{Bell-type correlations in many-body quantum systems have been investigated in several recent works \cite{Wang_2025,Schmied2016}. Specifically, in Section V of Ref. \cite{Wang_2025} Bell correlation depth is addressed, pointing to genuine multipartite nonlocality in a superconducting qubit system. In their work, the authors use the term "Bell-operator correlations" and focus on using system energy as a witness of Bell nonlocality for the Svetlichny operator associated with a GHZ state. This approach relies on a specific choice of measurement settings that simplify the Svetlichny operator (see \cite{Baccari_2019}), allowing the inequality to be tested with far fewer measurements. In contrast, our work takes a more direct path: we reconstruct all individual correlation terms involved in the Svetlichny inequality \eqref{e:SVL_N_Qubits} on quantum processors. This requires executing and measuring correlations across different, incompatible measurement contexts, and then evaluating the collective consistency of these measurements when estimating the expectation value of the Svetlichny operator. Our simulation is thus literal, emulating a physical test of the inequality on actual quantum hardware. While this approach does not constitute a loophole-free test, we demonstrate that—when combined with quantum random circuits—the protocol provides a powerful diagnostic tool. It yields rich information that enables a direct comparison between different hardware architectures, even detecting subtle variations in qubit behavior within a single processor.} 

\subsection{Noise Sensitivity of Multipartite Nonlocality Tests}
\label{s:Numerical}

\textcolor{red}{Before evaluating the protocol on real hardware, we use numerical simulations to compare the noise sensitivity of the Svetlichny and Mermin tests.} Our goal is to demonstrate that the nonlocality test based on Svetlichny's inequality is the hardest\footnote{When referring to Svetlichny as “harder”, we mean strictly its theoretical structure: the inequality detects GMNL that cannot be reduced to bilocal or hybrid-local models. Our simulations show a faster degradation of Svetlichny violations under realistic noise models, a behavior that is consistent with, though it does not by itself establish, the theoretical expectation that higher-order correlations tend to be more sensitive to experimental imperfections.}. \textcolor{blue}{As a reference, we include the Grover search~\cite{grover1996fast} performance as an example of how the nonlocality tests can quantify how well a QPU is working: good nonlocality test scores predict better algorithm performance. The choice of Grover is not essential; any algorithm with a well-defined performance metric, such as phase estimation~\cite{kitaev1995quantum} or the Bernstein-Vazirani algorithm~\cite{bernstein1997quantum}, would serve the same purpose. We consider the search for a single marked element in an unstructured database of $2^N$ elements with $N=3$, $4$, and $5$ qubits, as its success probability is a single scalar that decreases monotonically with noise and scales with qubit number without requiring ancillary registers.} We contrast the relative violation percentage of Svetlichny and Mermin under depolarizing noise, both normalized to the range $[0,1]$, against the success probability of Grover's algorithm under the same noise conditions. We find that at fixed noise the relative violation percentage for both inequalities decreases with increasing qubit number, and that this decrease is faster for the Svetlichny inequality, which supports our "hardness" assumption.

\subsection*{Simulation protocol}
\label{sec:protocol}

\begin{figure*}[h!]
    \centering
    \includegraphics[width=\linewidth]{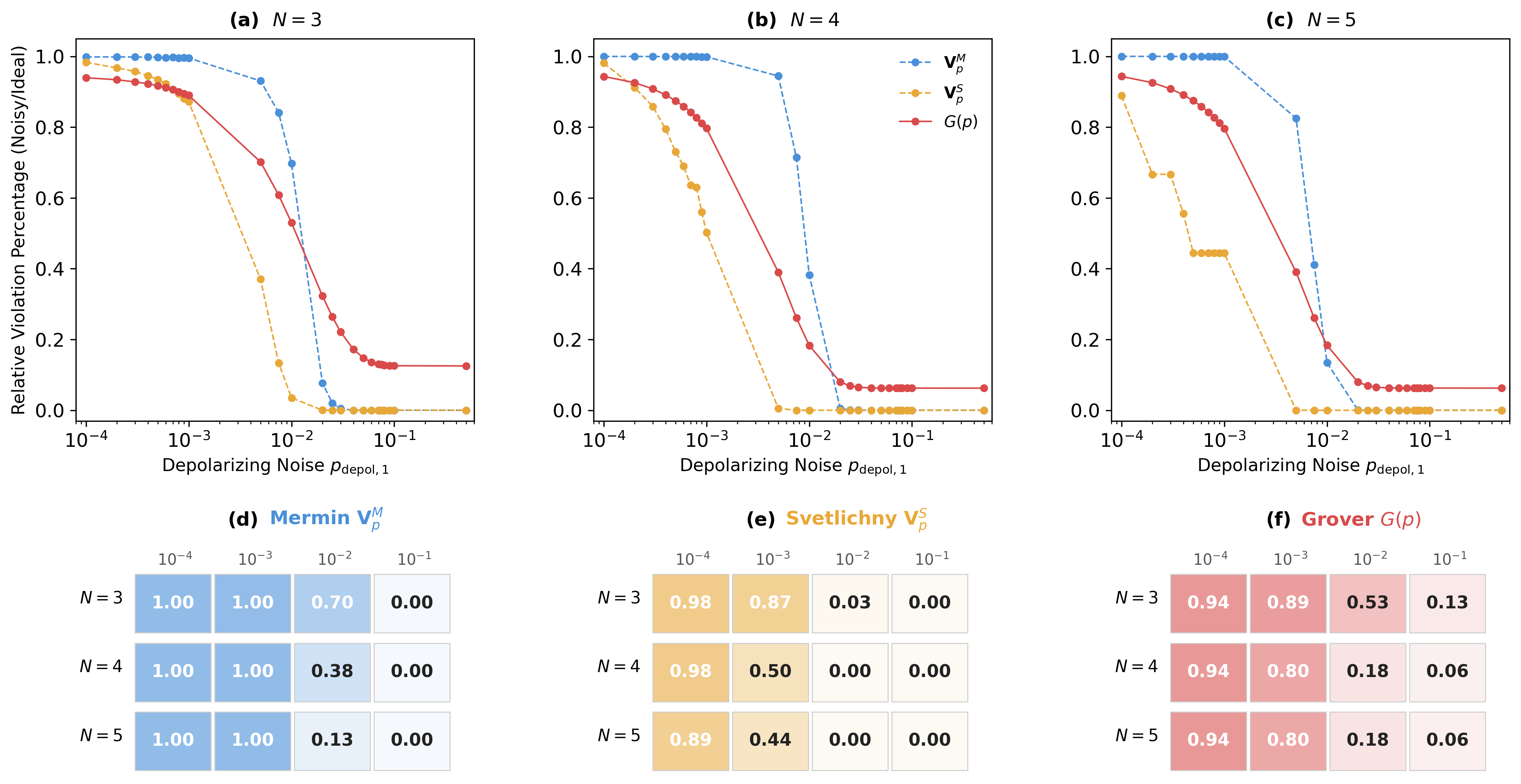}
    \caption{Comparison between the relative violation percentage of Svetlichny's inequality, that of Mermin's, and algorithmic success probability of Grover's algorithm for (a) $N=3$, (b) $N=4$, and (c) $N=5$ under varying one-qubit depolarizing noise $p_{\mathrm{depol},1}$ and two-qubit depolarizing noise $p_{\mathrm{depol},2}=2p_{\mathrm{depol},1}$. For this simulation, $10^4$ random circuits were generated, and the Svetlichny and Mermin inequalities were evaluated using $10^3$ shots. \textcolor{blue}{Panels (d)--(f) show heatmap summaries at representative noise levels ($p_{\mathrm{depol}} \in \{10^{-4}, 10^{-3}, 10^{-2}, 10^{-1}\}$); color intensity reflects the relative value (darker = closer to ideal noise-free case). }Note, the step-like behavior for $N=5$ arises due to the low number of violating states that decreases with increasing $N$ as depicted in~\cite{Granda_2025}.}
    \label{fig:grover_comparison}
\end{figure*}

The procedure for performing a nonlocality test is described in the steps below. The description refers to Svetlichny's inequality, but conducting Mermin's test is completely analogous (doing all the necessary changes \textit{mutatis mutandis}).

\begin{enumerate}
    \item For testing the nonlocality violation, we produce a dataset of $10^4$ quantum random circuits of a specific circuit depth. Each one of these circuits leaves the simulated quantum processor in a particular random (pure) state.
    \item 

    \textcolor{blue}{For each random circuit generated in the previous step, we numerically optimize the local measurement settings to maximize the Svetlichny operator expectation value $\langle S_N^{\pm} \rangle_\rho$. Each party $j$ performs one of two dichotomic measurements, defined by local spin observables of the form $\hat{n}_j \cdot \boldsymbol{\sigma}$, where $\hat{n}_j$ is a unit vector on the Bloch sphere and $\boldsymbol{\sigma} = (\sigma_x, \sigma_y, \sigma_z)$ is the vector of Pauli matrices. The optimization is performed over all $2N$ measurement directions $\{\hat{n}_j, \hat{n}'_j\}_{j=1}^N$. Note that the inequality itself is not modified: the Tsirelson bound $2^{N-1}\sqrt{2}$ remains fixed, but the measurement settings that maximize the observed violation depend on the quantum state prepared by each circuit and must be determined numerically.}
    \item Using the list $\mathbf{L}$ of maximal violation values, one for each random circuit, we create a histogram $\mathbf{H}_{I}$ of all obtained values corresponding to the ideal distribution. The histogram thus obtained will be referred to as \textit{ideal}. It represents the results expected from a noiseless quantum processor.
    \item As we have obtained the ideal distribution and percentage of violations for the dataset of random circuits, we fix a range of values of the depolarizing noise parameter $p$ for one qubit. The two-qubit depolarizing noise value is defined proportionally to the single-qubit depolarizing noise value. The same states and local measurements described in the above steps are run in a noisy quantum processor simulator with the same degree of depolarizing noise. The noisy violation values obtained will be stored in a list $\mathbf{L}_{p}$, where $p$ corresponds to the specified depolarizing noise for one and two qubits. Based on the values $\mathbf{L}_{p}$, we can visualize a noisy histogram $\mathbf{H}_{p}$. This procedure is repeated for different values of depolarizing noise $p$ in order to investigate the behavior under a change in~$p$.
    \item For each histogram, $\mathbf{H}_{I}$ or $\mathbf{H}_{p}$, one can compute the violation percentages, $\mathbf{V}_{I}$ and $\mathbf{V}_{p}$, which are the number of instances that produce a violation of the corresponding inequality. Naturally, the violation percentage tends to decrease as noise increases. Whenever necessary, in the interest of clarity, we add a superscript in order to distinguish magnitudes associated with Mermin or Svetlichny tests, i.e., we will use $\mathbf{V}^{M}_{p}$, $\mathbf{V}^{S}_{p}$, etc.
    \item In order to compare noisy and ideal results, one can compute the relative violation percentage, defined as $V_{r}(p)=\frac{\mathbf{V}_{p}}{\mathbf{V}_{I}}$. Notice that 
    $V_{r}(p)$ ranges from zero (corresponding to $p=\infty$) to one (corresponding to $p=0$). The magnitude $V_{r}(p)$ indicates the degree to which the simulated noisy quantum processor (here in the proof-of-principle scenario of depolarizing noise $p$) can reproduce the ideal results as a function of $p$.
    \item For completeness, we quantify the differences between the distributions $\mathbf{H}_{I}$ and $\mathbf{H}_{p}$ in terms of distribution change quantifiers, such as the Maximum Mean Discrepancy $\text{MMD}(\mathbf{H}_{I},\mathbf{H}_{p})$. 
\end{enumerate}
We compare the relative violation percentages $V_{r}^M(p)$ and
$V_{r}^S(p)$ with the success probability of Grover's
algorithm~\cite{grover1996fast} under the same noise conditions.
For Grover's algorithm, we consider the search for a unique marked
element in a database of $2^N$ elements. The success probability
$G(p)$ is defined as the fraction of measurement shots that return
the correct marked element under depolarizing noise~$p$.

\subsection*{Numerical analysis}
\label{sec:numerical_analysis}

In this section, we present the results obtained following the protocol of the nonlocality test described above, and we aim to compare the success probability of Grover's algorithm to nonlocality tests under varying noise parameters. This analysis serves as an important first step before evaluating on real hardware.

\textcolor{blue}{In Fig.~\ref{fig:grover_comparison} panels (a), (b), and (c)} we plot the relative violation percentage of Mermin and Svetlichny histograms for three, four and five qubits with respect to the depolarizing noise parameter $p$, i.e., the values $V^{M}_{r}(p)$ and $V^{S}_{r}(p)$. As we add depolarizing noise for one and two qubits, we set the two-qubit depolarizing noise to twice the depolarizing noise applied to one qubit. Those results are presented together with the success probability of Grover's algorithm~$G(p)$ of finding a unique marked state. In Fig.~\ref{fig:grover_comparison}, all three panels (a), (b), and (c), clearly, $\mathbf{V}^{S}_{p}$ drops off faster than $\mathbf{V}^{M}_{p}$ and~$G(p)$. This indicates that the histograms associated with the Svetlichny inequality are more sensitive to noise than those associated with Mermin inequalities and the success probability of Grover. 
\textcolor{blue}{Consequently, under the same noise conditions, the Svetlichny violation fraction degrades faster than both the Mermin violation fraction and Grover's success probability}. For five qubits, as seen in Fig.~\ref{fig:grover_comparison}(c), the Svetlichny violation fraction exhibits a noise dependence that more closely tracks the degradation observed in Grover's success probability than the Mermin violation fraction does.

While the relative violation percentage captures the fraction of circuits that remain above the violation threshold, it does not quantify how the full distribution of Bell operator expectation values changes under noise. To address this, we also evaluate the Maximum Mean Discrepancy (MMD)~\cite{gretton2012kernel} in Fig.~\ref{fig:mmd_rbf_all} for various qubit numbers $N=3,4,$ and $5$. The MMD is defined as a two-sample test~\cite{gretton2012kernel} where the sample distance of samples $x$ and $y$ is compared in the respective kernel feature space:

\begin{figure}[H]
    \centering
    \includegraphics[width=\columnwidth]{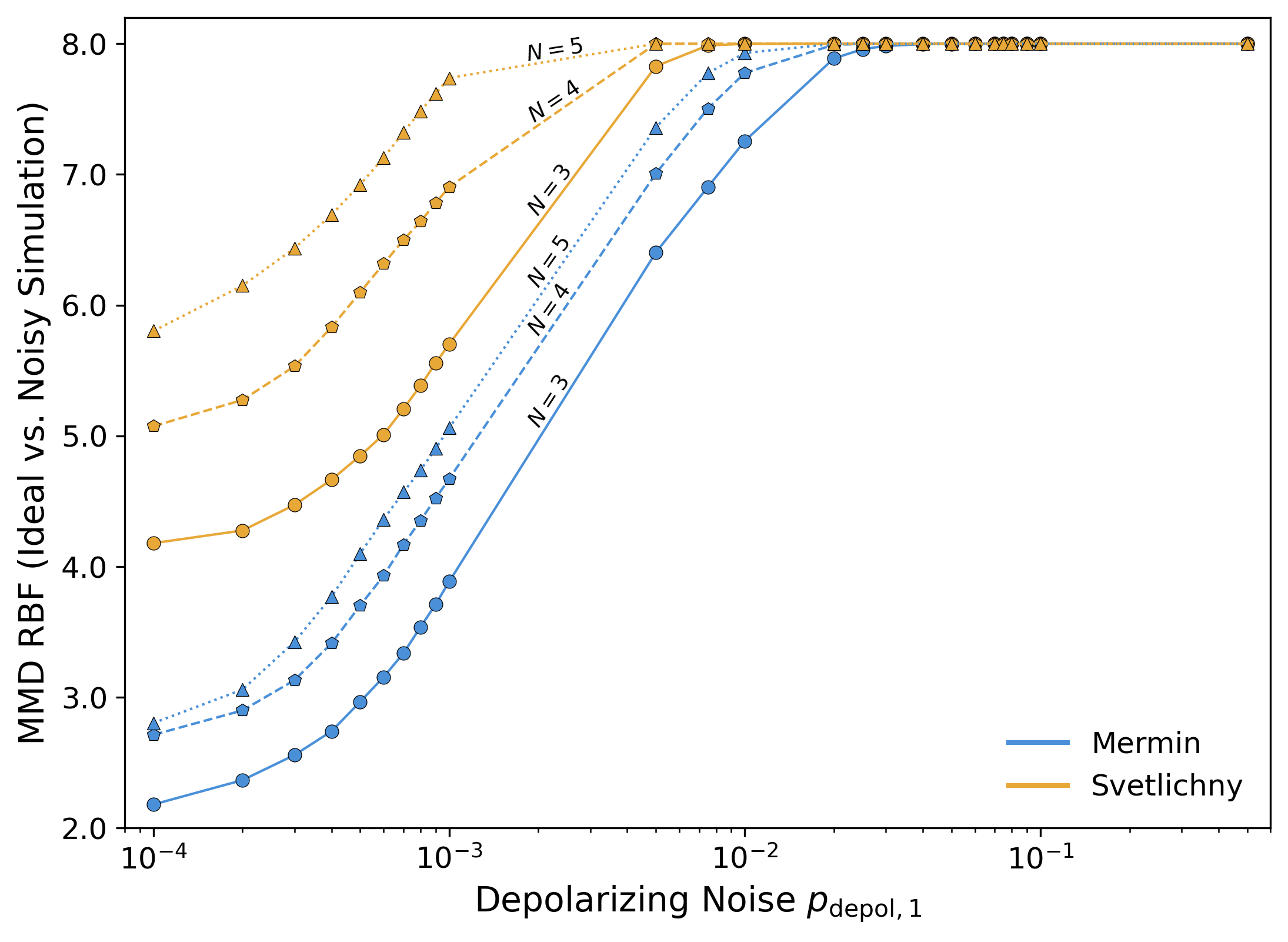}
    \caption{Evaluating the MMD with a Gaussian kernel as defined in \textcolor{blue}{Eq.}~\eqref{eqn:mmd}  between the ideal (simulated) and noisy violation distributions  for $N = 3, 4, 5$ qubits based on $10^4$ samples while varying over the depolarizing noise for one and two qubits, respectively. The simulation is using $10^3$ shots.}
    \label{fig:mmd_rbf_all}
\end{figure}

\begin{equation}
\label{eqn:mmd}
\begin{aligned}
\mathrm{MMD}^2(p,q) :=\;&
\mathbb{E}_{x \sim p,\, y \sim p}\!\left[k(x,y)\right]
- 2\,\mathbb{E}_{x \sim p,\, y \sim q}\!\left[k(x,y)\right] +
\\
&\mathbb{E}_{x \sim q,\, y \sim q}\!\left[k(x,y)\right]
\end{aligned}
\end{equation}
with the Gaussian kernel defined as:
\begin{align}
    k(x,y) = \exp \left( - \frac{(x-y)^2}{2 \sigma}\right)
\end{align}
and $\sigma \in [0.01, 0.1, 1, 10, 100, 1000]$ being the bandwidth parameter. \textcolor{blue}{We refer to Eq.}~\eqref{eqn:mmd} simply as MMD. This evaluation also demonstrates the higher sensitivity of Svetlichny violations to noise than Mermin violations exhibit, as the difference in distribution increases faster.
Based on these simulations, we find that violations of the Svetlichny
inequality are more sensitive to depolarizing noise than violations of the Mermin inequality. This behavior is consistent with the fact that Svetlichny inequalities are designed to rule out all HLR models and therefore probe a stronger form of multipartite correlation. These observations support the use of Svetlichny violations as a stringent
benchmark for assessing the collective performance of quantum
processors. In the following sections, we examine both tests on the Marmot and Garnet quantum processors, highlighting their applications for calibration purposes.


\section{Experimental Methodology and Device Evaluation}
\label{s:Experiments}

\begin{table*}[htbp]
\centering
\small
\caption{Comparison of selected hardware characteristics of the Garnet and Marmot quantum processors. The values are taken from the references cited in the main text and from the benchmark tests reported by Piskor \emph{et al.}~\cite{psb25}.}
\label{tab:quantum-hardware}

\begin{tabular}{lcc}
\toprule
\textbf{Property} & \textbf{Garnet} & \textbf{Marmot} \\
\midrule

Physical qubits
& 20
& 16 (12) \\

Single-qubit gate fidelity (\%)
& mean: 99.85; median: 99.89
& $99.96\pm0.01$ \\

Two-qubit gate fidelity (\%)
& mean: 98.59; median: 99.06
& $98.47\pm0.30$ \\

Connectivity / topology
& Square lattice with tunable couplers
& All-to-all \\

Quantum volume
& 32
& 128 \\

\bottomrule
\end{tabular}

\end{table*}

\textcolor{red}{Building on the numerical analysis of Section~\ref{s:Numerical}, for the Mermin and Svetlichny inequalities, $10^5$ random quantum 
circuits were generated for each of the families described above 
(Clifford and Clifford with the \texttt{T} gate) and each qubit 
number ($N = 3, 4, 5$), using a universal gate set (\texttt{H}, 
\texttt{S}, \texttt{CNOT}, \texttt{T}, \texttt{Rx}, \texttt{Rz}); 
the use of native gates is examined separately in 
Subsection~\ref{ss:Native_Gates}. For GHZ states, standard 
preparation circuits were adopted. The resulting histograms describe 
how the states are distributed across different ranges of violation values.}

\textcolor{red}{Given the high computational cost of experimentally testing each circuit, the most representative ones were selected across the full nonclassical region~\cite{Granda_2025}: it was divided into five equal sub-intervals, and from each one the circuit with the highest 
violation value, lowest depth, and fewest entangling (\texttt{CNOT}) gates was selected\footnote{For the initial generation of theoretical histograms, random circuit depths between 55 and 75 layers were used. However, in the 4- and 5-qubit cases, the percentage of circuits that violated the Svetlichny inequality was already extremely small, even in the ideal case without any noise model applied~\cite{Granda_2025}.  Reducing the depth to the range 3--15 layers produced states distributed across the entire nonclassical interval, allowing the identification of circuits capable of achieving significant violations at sufficiently low depth.}. This 
strategy was applied only to the Clifford with the \texttt{T} gate 
set. For the Clifford group, due to its discrete structure and its 
correspondence with stabilizer states, no subdivision of the region 
was required; only the depth and CNOT-count criteria were applied.}

\begin{figure}[!h]
    \centering
    \includegraphics[width=\columnwidth]{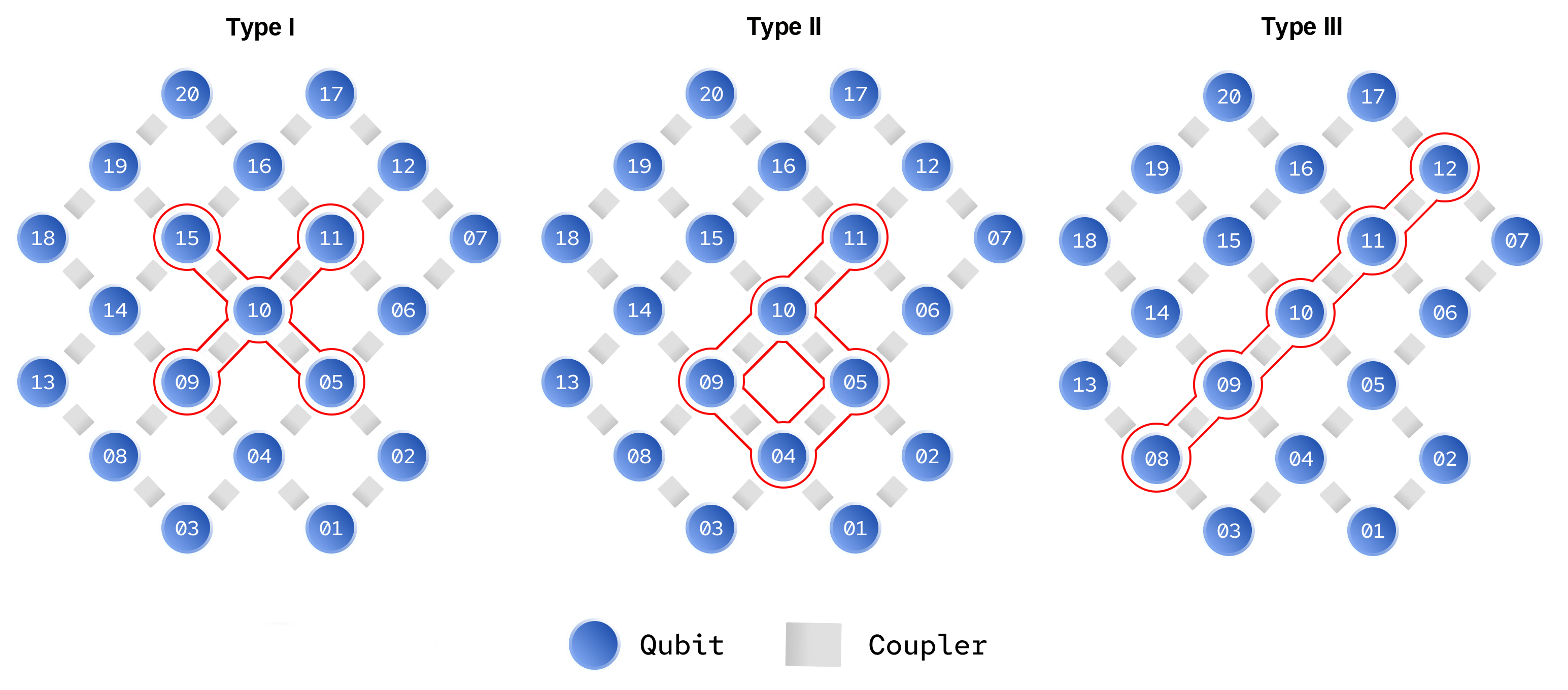}
    \caption{Examples of connectivity configurations for a five-qubit system. Connectivity layout adapted from IQM (Garnet QPU)~\cite{IQM_Garnet}. Highlighted qubit subsets and red connections added by the authors.}
    \label{fig:5-qubit_conectivity}
\end{figure}

\textcolor{red}{Garnet has a fixed-coupling square-lattice architecture that 
requires SWAP-based routing for non-adjacent qubits, whereas Marmot 
provides all-to-all connectivity (see 
Table~\ref{tab:quantum-hardware} for a comparison of their 
main hardware characteristics). Given this topological difference, 
the same logical circuits were executed on both platforms, so that 
the comparison reflects the full-stack performance of each processor 
(hardware architecture and compilation layer). 
Fig.~\ref{fig:5-qubit_conectivity} shows examples of five-qubit 
connectivity subgraphs on the Garnet lattice; different topologies 
produce different compiled circuits, which affects the achievable 
violation values for logically identical circuits.}

\textcolor{red}{The number of shots per circuit was set to 2000 for Garnet and 200 for Marmot, the maximum allowed per job. Two calibrated noise model simulators were also used: the \texttt{FakeAdonis} backend simulates the calibrated noise model of Garnet, and the \texttt{AQT Provider} simulator that of Marmot. Comparing real executions with their simulated counterparts, we evaluate how well the reported noise models reproduce the observed behavior and identify deviations from 
non-modeled effects.}

\textcolor{red}{Since the evaluation of Bell-type inequalities is sensitive to sample size, this difference must be taken into account when 
interpreting the results. A complementary stability analysis was performed to quantify the effect of shot noise, particularly relevant for Marmot (Appendix~\ref{appendix:shots}). The 
quantitative metrics used to compare both devices are defined in Appendix~\ref{appendix:metrics}. We also investigate how simulated nonlocal correlations depend on the specific physical qubits used, by replicating the same circuit across different qubit subsets within each device (see Subsections~\ref{ss:Triplets} and~\ref{ss:Native_Gates}).}

\subsection{GHZ States as a Reference for Multipartite Nonlocality}
\label{ss:GHZ_States}

\begin{figure*}[h!]
\centering
\includegraphics[width=\textwidth]{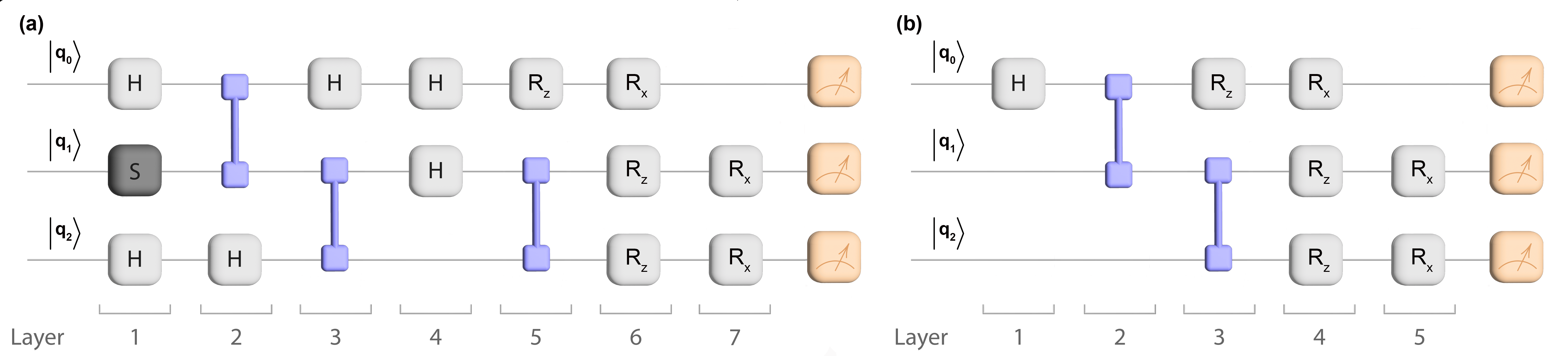}
\caption{\textcolor{blue}{Quantum circuits for $N=3$ qubits with the same theoretical Svetlichny violation ($4\sqrt{2}$). (a) Clifford circuit (7 layers, \texttt{H}, \texttt{S} gates). (b) GHZ preparation (5 layers). Both include measurement basis rotation gates (\texttt{Rz}, \texttt{Rx}). The greater depth 
of the Clifford circuit results in more noise accumulation, which accounts for the difference in measured violation 
values.}}
\label{fig:clifford_vs_GHZ}
\end{figure*}

\begin{figure*}[!htb]
\centering
\includegraphics[width=0.96\textwidth]{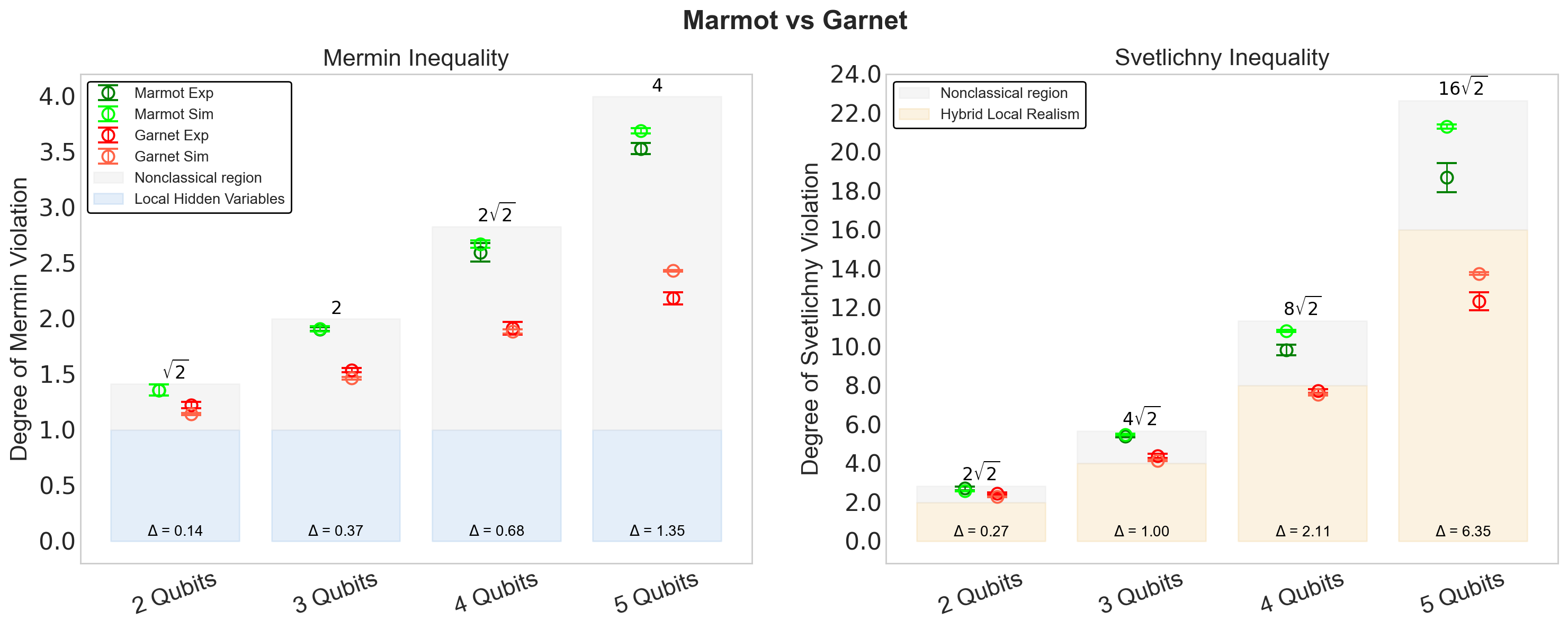}
\caption{Comparison between Marmot and Garnet for GHZ states with 3--5 qubits. The left panel shows the violation of 
Mermin's inequality, and the right panel shows the violation 
of Svetlichny's inequality. Each data point corresponds to the mean value from 20 experimental repetitions using 2000 shots for Garnet and 200 shots for Marmot, \textcolor{blue}{with error bars indicating one standard deviation ($\pm \sigma$)}. The labels \textit{Garnet Sim} and \textit{Marmot Sim} refer to the 
noise-model simulators \texttt{FakeAdonis} and 
\texttt{AQT-Provider}, respectively. Shaded regions indicate the classical bound of each inequality: Local Hidden Variables (blue) and Hybrid Local Realism (gold); the grey region above each bound is the nonclassical region. 
The parameter $\Delta$, shown at the bottom of each group, represents the absolute deviation between the mean experimental violation values measured on Marmot and Garnet, as defined in Eq.~\eqref{eq:delta_plat}. For completeness, we include the corresponding values for the two-qubit CHSH 
case.}

\label{fig:GHZ_IQM_MERMIN_SVET}
\end{figure*}

\begin{table*}[htbp]
\centering
\small
\caption{Theoretical, experimental, and simulated results for the Mermin and Svetlichny inequalities using GHZ states from 3 to 5 qubits. For Garnet, both the experimental data and the \texttt{FakeAdonis} simulator outputs are reported. For Marmot, results from the Marmot processor and its simulator (\texttt{AQT-Provider}) are included. Mean values and standard deviations ($\mu \pm \sigma$) are computed from 20 experimental repetitions, together with the metrics $\varepsilon_{\mathrm{rel}}$ and $R$. Negative values of $R$ indicate results below the corresponding HLR bound. Values for the two-qubit CHSH case (normalized) are included for completeness.}
\label{tab:GHZ_IQM_Mermin_Svetlichny}

\begin{tabular}{lcccccccccc}
\toprule

& & &
\multicolumn{2}{c}{\textbf{Garnet ($\mu\pm\sigma$)}} &
\multicolumn{2}{c}{\textbf{Marmot ($\mu\pm\sigma$)}} &
\multicolumn{2}{c}{\textbf{Garnet}} &
\multicolumn{2}{c}{\textbf{Marmot}} \\

\cmidrule(lr){4-5}
\cmidrule(lr){6-7}
\cmidrule(lr){8-9}
\cmidrule(lr){10-11}

\textbf{Inequality}
&\textbf{$N$} & \textbf{Theor.} & \textbf{Exp.} & \textbf{Sim.} & \textbf{Exp.}
&\textbf{Sim.}&$\boldsymbol{\varepsilon_{\mathrm{rel}}}$&$\boldsymbol{R}$&
$\boldsymbol{\varepsilon_{\mathrm{rel}}}$&$\boldsymbol{R}$
\\

\midrule

CHSH&2&1.41&1.22 $\pm$ 0.03
&1.14 $\pm$ 0.01&1.36 $\pm$ 0.10&1.29 $\pm$ 0.02&13.5&0.54&3.9&0.87
\\

\addlinespace
\midrule

\multirow{3}{*}{Mermin}
&3&2&1.54 $\pm$ 0.02&1.46 $\pm$ 0.01&1.90 $\pm$ 0.01&1.90 $\pm$ 0.02&23.0&0.54&4.7&0.91
\\

&4&2.82&1.91 $\pm$ 0.06&1.88 $\pm$ 0.02&
2.59 $\pm$ 0.08&2.67 $\pm$ 0.03&32.3&0.50&8.2&0.87
\\

&
5 & 4 & 2.18 $\pm$ 0.05 & 2.43 $\pm$ 0.01 & 3.52 $\pm$ 0.04 & 3.68 $\pm$ 0.02 & 45.4 & 0.39 & 11.8
& 0.84
\\

\addlinespace
\midrule

\multirow{3}{*}{Svetlichny}
& 3 & 5.65 & 4.39 $\pm$ 0.10 & 4.15 $\pm$ 0.03 & 5.39 $\pm$ 0.05 & 5.48 $\pm$ 0.04 & 22.3 & 0.24 & 4.6 & 0.84
\\

&4&11.31&7.72 $\pm$ 0.09&7.54 $\pm$ 0.06&
9.83 $\pm$ 0.26&10.81 $\pm$ 0.07&31.7&-0.08&13.1&0.55
\\

&
5
&
22.62
&
12.3 $\pm$ 0.50
&
13.8 $\pm$ 0.07
&
18.68 $\pm$ 0.74
&
21.30 $\pm$ 0.10
&
45.5
&
-0.55
&
17.4
&
0.40
\\

\bottomrule
\end{tabular}

\end{table*}

With the methodological framework established, we begin by testing the quantum devices using the GHZ state. This choice provides a natural starting point for benchmarking, as it is the maximally entangled state, produces the strongest theoretical violations for the inequalities considered, and can be prepared as a very compact, low-depth circuit (\textcolor{blue}{see Fig.~\ref{fig:clifford_vs_GHZ} for a comparison 
with a representative Clifford circuit of equal 
theoretical violation}). This provides a clean reference for evaluating the devices across the remaining circuit families.

A scalability analysis was carried out for systems of 2 to 5 qubits using the Mermin and Svetlichny inequalities on the Garnet and Marmot quantum computers. This allows us to study initially the bipartite or extended correlations (captured by Mermin) and then \textcolor{blue}{simulated} genuine multipartite correlations (detected by Svetlichny), and to corroborate how increasing the number of qubits affects the device's ability to sustain such correlations.

Fig.~\ref{fig:GHZ_IQM_MERMIN_SVET} shows the experimental results and those from the noise-model simulators, where each point represents the mean violation value together with one standard deviation. For reference, the corresponding ideal theoretical values and the bounds imposed by the LHV and HLR models are also shown. This allows a direct visualization of how the violation changes as the number of qubits increases and how both devices perform under the same reference circuits.

\textcolor{red}{To establish a comparison between both devices, we use: the mean relative error $\varepsilon_{\mathrm{rel}}$, which quantifies the percentage deviation between the experimental value and its corresponding theoretical prediction; the normalized violation ratio $R(S_i,T_j)$, which quantifies the relative performance of a given experimental configuration by normalizing the measured violation between the classical and quantum bounds; and the parameter $\Delta$, which quantifies the absolute deviation between the mean violation values obtained on both devices (for further details see Appendix~\ref{appendix:metrics}).}

Table~\ref{tab:GHZ_IQM_Mermin_Svetlichny} shows that the Svetlichny inequality yields negative values of $R(S_i,T_j)$ for the Garnet processor starting from four qubits ($R=-0.08$ for $N=4$ and $R=-0.55$ for $N=5$). This means that, in these cases, the experimental value is very close to the classical bound, or even falls below it, indicating that the IQM processor can no longer \textcolor{blue}{simulate} the type of correlations required for GMNL. In contrast, all results for the Mermin inequality remain within the \textcolor{blue}{nonclassical} region for $2 \le N \le 5$, showing that the device can still \textcolor{blue}{simulate} correlations that are nonlocal in the weaker sense captured by Mermin's inequality.

The Svetlichny inequality is more demanding because it depends on higher-order correlations. As a result, errors induced by decoherence, imperfect two-qubit gates, or loss of phase coherence accumulate and become amplified; this can suppress the violation even when the relative error affecting each individual term is similar to that observed in the Mermin case. This indicates that, for $N>3$, the noise affecting the processor degrades genuine multipartite correlations more strongly than bipartite or extended correlations, making Svetlichny violations particularly fragile under realistic experimental conditions.

From a benchmarking perspective, these results position Marmot as the processor exhibiting a higher value of $R$ and a lower mean relative error $\varepsilon_{\mathrm{rel}}$.

\subsection{CLIFFORD + \texttt{T} GROUP}
\label{ss:Clifford+T}

\textcolor{red}{The analysis for the Clifford with the \texttt{T} gate group is consistent with the experimental selection criteria described previously (one representative circuit per sub-interval with the highest violation value while keeping the logical description as simple as possible: lowest depth and fewest entangling gates). Adding the \texttt{T} gate to the Clifford set turns the system into a universal quantum gate set, where the distribution of violation values no longer concentrates around a small set of discrete outputs, as occurs with Clifford-only circuits, but instead spreads over a continuous range within the nonclassical region, allowing the circuits to reach violation values that are otherwise not accessible.}

\begin{figure*}[h]
\centering
\includegraphics[width=\textwidth]{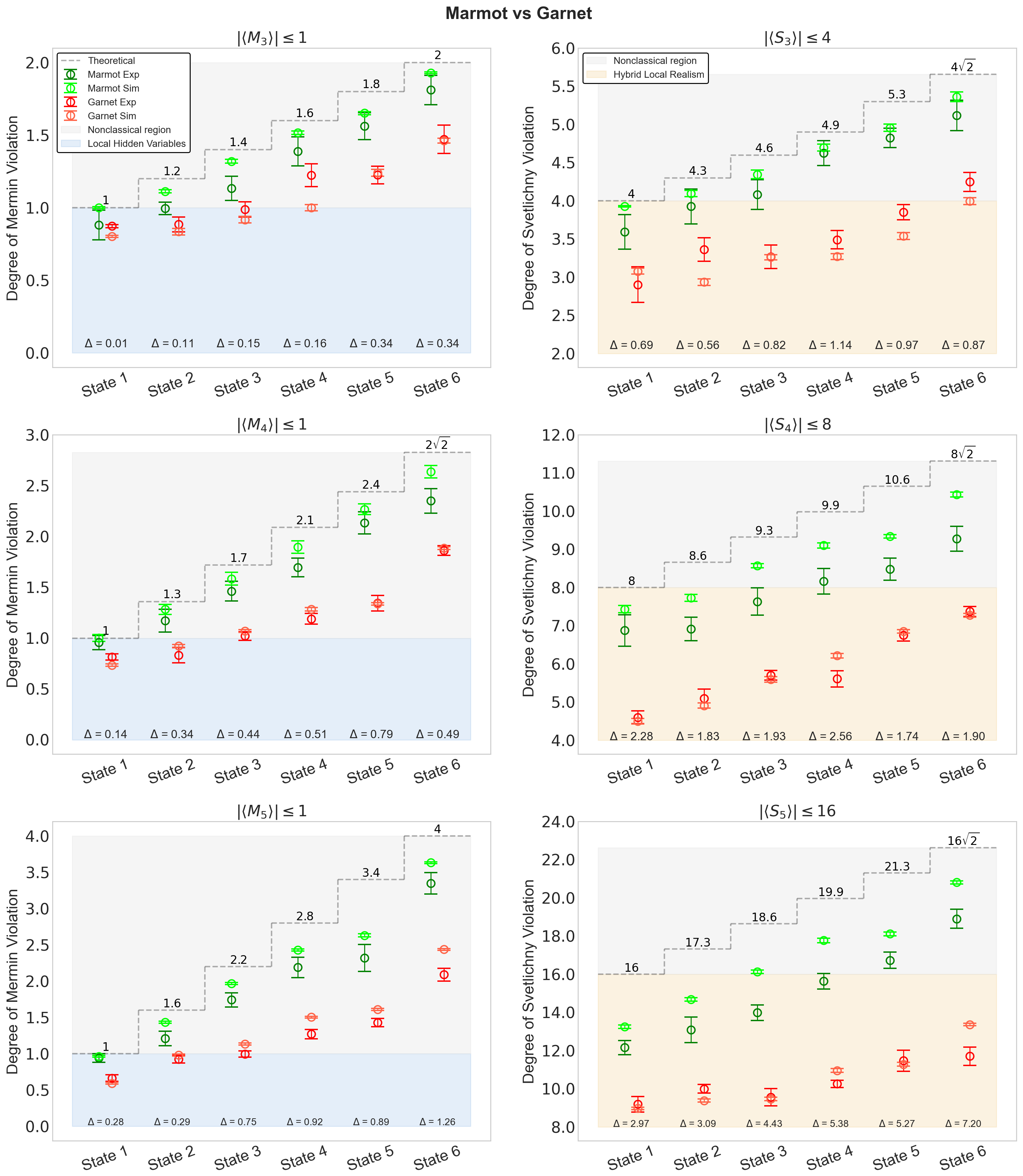}
\caption{Comparison between Garnet and Marmot for Clifford with the \texttt{T} gate circuits with (a) 3 qubits, (b) 4 
qubits, and (c) 5 qubits. The left panels show Mermin's inequality and the right panels Svetlichny's inequality. 
Each point corresponds to the mean value from 20 experimental repetitions using 2000 shots for Garnet and 200 shots for Marmot, \textcolor{blue}{with error bars indicating one standard deviation ($\pm \sigma$)}. The labels \textit{Garnet Sim} and 
\textit{Marmot Sim} refer to the noise-model simulators 
\texttt{FakeAdonis} and \texttt{AQT-Provider}, respectively. Shaded regions indicate the classical bound of each 
inequality: Local Hidden Variables (blue) and Hybrid Local Realism (gold); the grey region above each bound is the nonclassical region. The parameter $\Delta$, shown at the bottom of each state, represents the absolute 
deviation between the mean experimental violation values measured on Marmot and Garnet.}
\label{fig:ALL_CLIFFORD+T_IQM_AQT}
\end{figure*}

The results are presented as a simultaneous comparison between the Mermin and Svetlichny inequalities for both devices (Garnet and Marmot). Fig.~\ref{fig:ALL_CLIFFORD+T_IQM_AQT} shows, for each selected representative state, the experimental violation values together with the predictions from their respective noise-model simulators and the corresponding theoretical bounds.

The results obtained are consistent with the general behavior observed for GHZ states: the Svetlichny inequality is violated only in the last sub-interval (State 6) for three qubits, while for four and five qubits no violation is observed anywhere along the nonclassical region. Mermin violations, in contrast, persist for all $N$ considered.

The quantitative results are presented in Table~\ref{tab:CliffordT_results} (Appendix~\ref{appendix:Tables}), where it can be observed that as the number of qubits increases, noise and imperfections accumulate, causing the multipartite correlations required by the Svetlichny inequality to become too weak to be detected under the experimental conditions available on the device.

\subsection{CLIFFORD GROUP}
\label{ss:Clifford}

\textcolor{red}{This section is dedicated to the structure and behavior of circuits constructed exclusively from the Clifford group, which are classically simulable, nonuniversal, and directly linked to the algebraic properties of stabilizer states. This group is nevertheless capable of producing states that maximally violate multipartite Bell inequalities, as do GHZ states. However, the distribution of violation values it generates exhibits a discrete structure, in contrast to the quasi-continuous distributions obtained with Clifford with the \texttt{T} gate circuits or with Haar-random unitary states.}

Rather than dividing the nonclassical region into sub-intervals, we directly selected the set of values that emerges from the theory (corresponding to the representative states $S_i$ defined in Appendix~\ref{appendix:metrics}), namely those capturing the characteristic violation levels associated with the different stabilizer structures generated by Clifford circuits.

The results are presented in Fig.~\ref{fig:ALL_CLIFFORD_IQM_AQT} (see also Table~\ref{tab:Clifford_results} in Appendix~\ref{appendix:Tables}). As in the previous sections, the behavior observed here follows the same general patterns. In particular, the Svetlichny inequality on the Garnet device ceases to exhibit violation for $N \ge 4$, indicating that it is unable to simulate the genuine multipartite correlations for larger qubit numbers. In contrast, the Mermin inequality continues to exhibit the ability to simulate violations across all cases considered.

It is well known that the largest violation within the discrete set of values generated by the Clifford group (independently of the number of qubits) corresponds to states that share the same theoretical violation value as GHZ states; in this sense, they can be viewed as a related subfamily from a theoretical perspective. However, the experimental violation values obtained for these Clifford circuits are systematically lower than those observed for GHZ. This difference does not arise from theoretical properties of the states but from practical aspects of their physical implementation.

As shown in Fig.~\ref{fig:clifford_vs_GHZ}, the latter are prepared using extremely compact circuits that rely only on \texttt{H} and \texttt{CNOT} gates, whereas the Clifford circuits used here require additional layers of \texttt{H}, \texttt{S}, \texttt{CNOT}, \texttt{Rx}, and \texttt{Rz} gates applied in random configurations. These additional layers and gates make them more susceptible to decoherence and gate errors. As a result, the experimental values obtained are lower, even though the theoretical prediction for both families is the same.

As shown in Table~\ref{tab:Clifford_results} (Appendix~\ref{appendix:Tables}), the relative error values ($\varepsilon_{\mathrm{rel}}$) obtained for the Clifford group fall within the same range as those observed for GHZ and Clifford with the \texttt{T} gate states. This quantitative comparison confirms that the reduction in violation values follows a comparable trend across the different families of states, without revealing any qualitatively distinct behavior specific to the Clifford case.

\onecolumn

\begin{figure*}[h!]
\centering
\includegraphics[width=\textwidth]{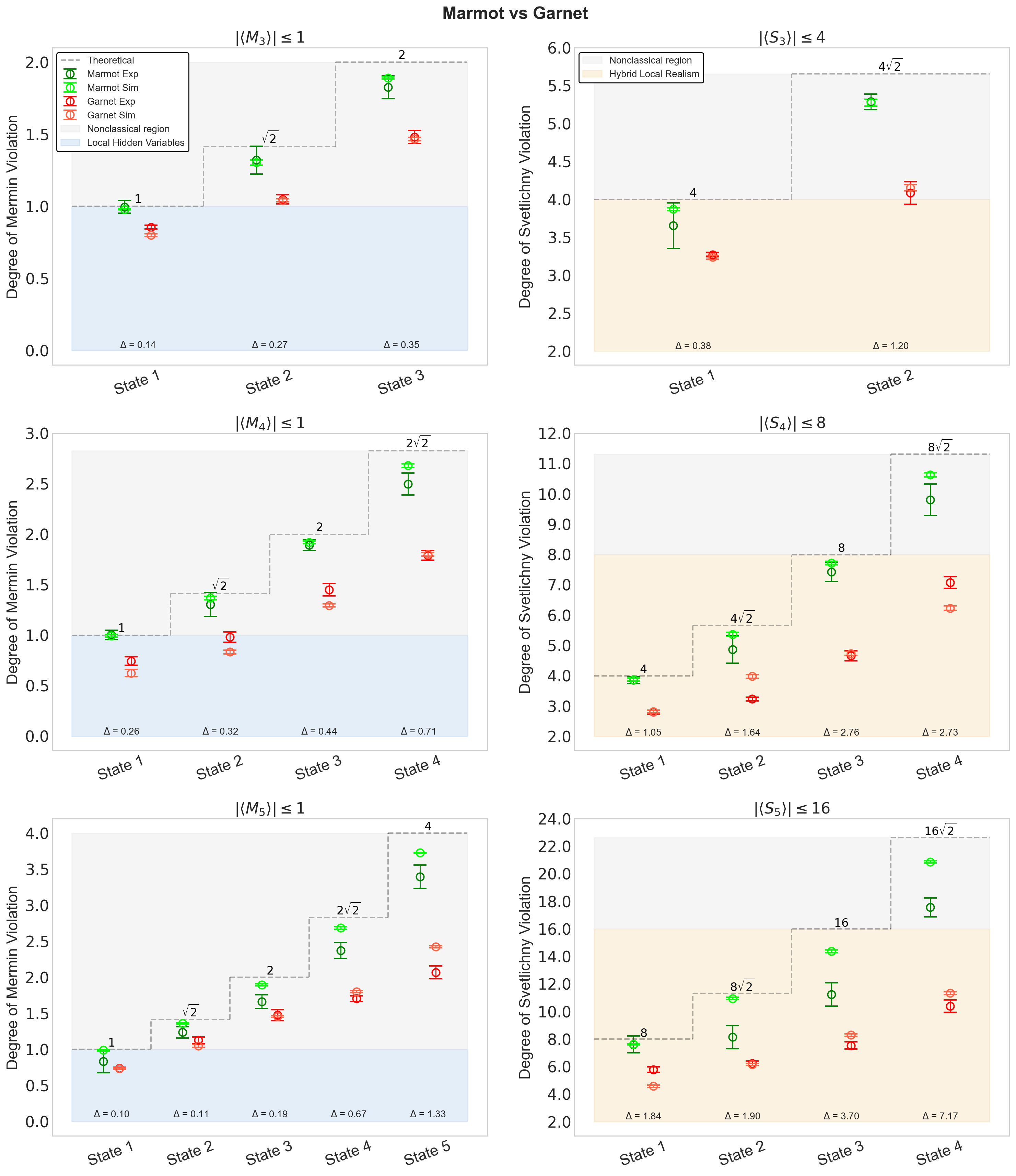}
\caption{Same as Fig.~\ref{fig:ALL_CLIFFORD+T_IQM_AQT}, but for Clifford circuits with (a) 3 qubits, (b) 4 qubits, and (c) 5 qubits. The left panels show Mermin's inequality and the right panels Svetlichny's inequality. Each point 
corresponds to the mean value from 20 experimental repetitions using 2000 shots for Garnet and 200 shots for 
Marmot, \textcolor{blue}{with error bars indicating one standard deviation ($\pm \sigma$)}. The labels \textit{Garnet Sim} and \textit{Marmot Sim} refer to the noise-model simulators \texttt{FakeAdonis} and \texttt{AQT-Provider}, respectively. Shaded regions indicate the classical bound of each 
inequality: Local Hidden Variables (blue) and Hybrid 
Local Realism (gold); the grey region above each bound is the nonclassical region. The parameter $\Delta$, 
shown at the bottom of each state, represents the absolute deviation between the mean measured violation values on Marmot and Garnet.}
\label{fig:ALL_CLIFFORD_IQM_AQT}
\end{figure*}

\twocolumn
\subsection{Impact of qubit location on simulated genuine multipartite nonlocality}
\label{ss:Triplets}

For Marmot, since ions are fundamentally identical, one might expect uniform performance across the various qubit subsets of the processor. This is an advantageous point for ion-trap quantum processors with regard to other architectures. Nevertheless, variations may arise due to boundary effects; for instance, an ion at the edge of the crystal has only a single neighbor, which could influence its coherence or gate behavior. \textcolor{red}{In this section, we test this hypothesis by executing identical circuits on different qubit subsets}.

\begin{figure}[h!]
\centering
\includegraphics[width=\columnwidth]{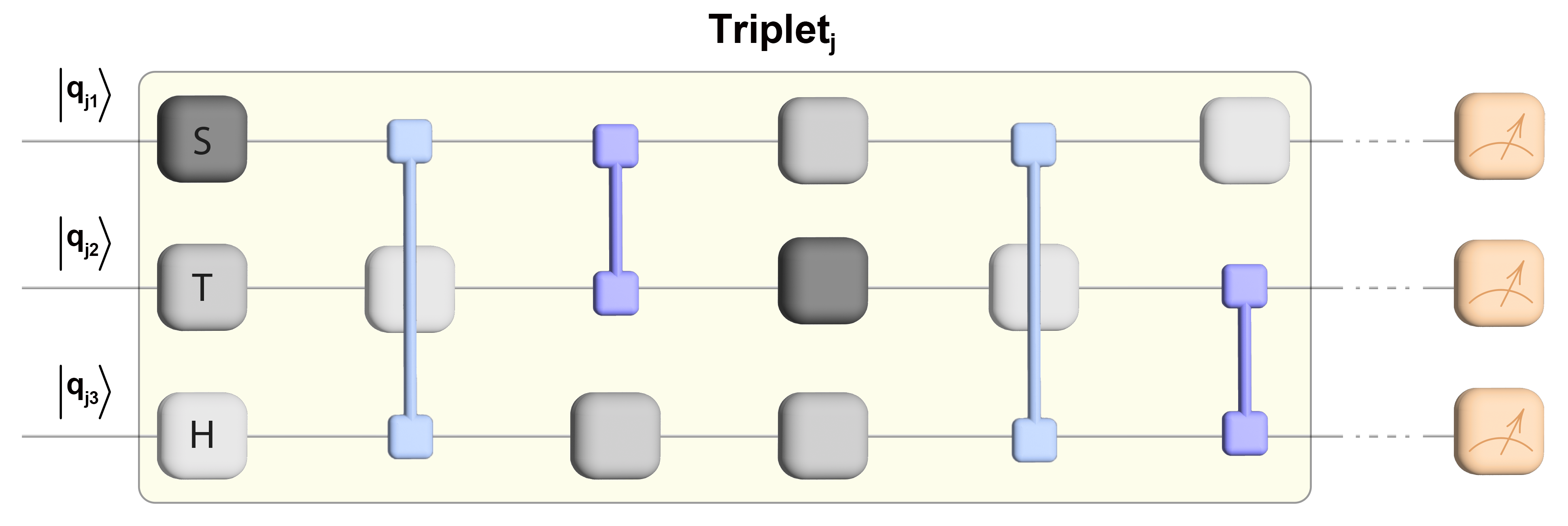}
\caption{
Schematic representation of the quantum circuit applied to a generic triplet of qubits  $T_j$  within the processor.
The same circuit is independently executed on different triplets selected across the device to test possible spatial variations in performance.}
\label{fig:Tripets_esquema}
\end{figure}

\textcolor{red}{We designed an experiment using all available qubits of the processor, repeating the same randomly generated three-qubit circuit on different qubit triplets across the full register (a schematic representation of this experiment is shown in Fig.~\ref{fig:Tripets_esquema})}. At the time the experiments were performed, twelve qubits were available in total (4 triplets). In this way, we test the behavior of Marmot under stress. As we show below, this is reflected in a higher dispersion of the mean values obtained for the Svetlichny operator (i.e., the left-hand side of Eq.~\eqref{e:SVL_N_Qubits}).

\begin{figure}[h!]
\centering
\includegraphics[width=\columnwidth]{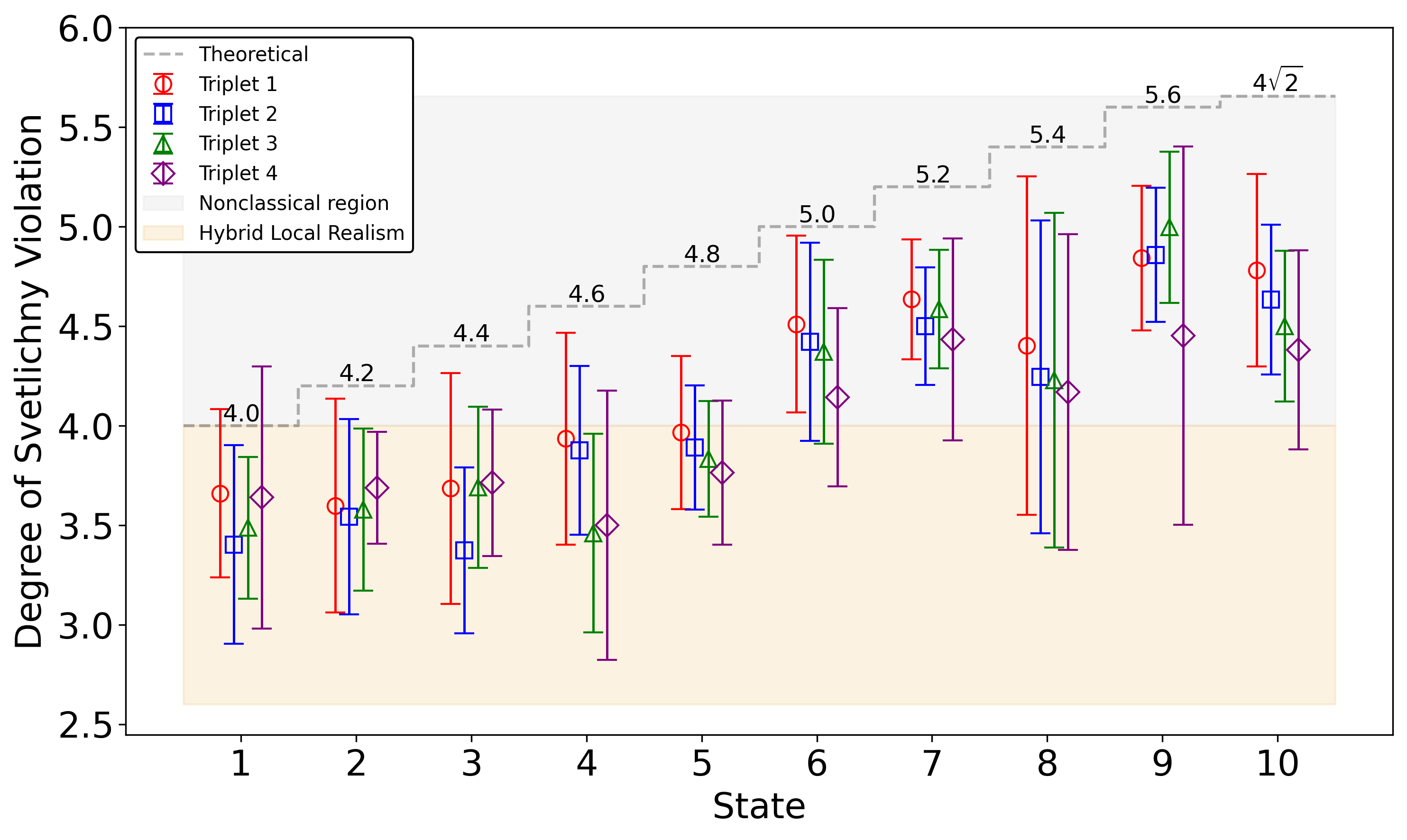}
\caption{Comparison of the Svetlichny inequality violations 
measured for four different qubit triplets within the Clifford 
with the \texttt{T} gate family. Each data point corresponds 
to the mean value over 20 experimental repetitions using 200 
shots for Marmot, \textcolor{blue}{with error bars indicating 
one standard deviation ($\pm \sigma$)}. Shaded regions 
indicate Hybrid Local Realism (gold) and the nonclassical 
region (grey). This analysis shows how the choice of physical 
qubits affects the measured violations on an ion-trap 
processor. Since the full set of available qubits is used, 
the dispersion across repetitions is larger than in 
Fig.~\ref{fig:ALL_CLIFFORD+T_IQM_AQT}. Notice also that the bars might depend on the chosen test and triplet (as is the case in state 9 in the plot).}
\label{fig: 3-qubit_Svetlichny_Triplet_1-2-3}
\end{figure}

\textcolor{red}{To ensure homogeneous calibration conditions, the full set of triplets was submitted within the same job; this was repeated 20 times to compute the corresponding mean values and standard deviations (to account for possible variations in calibration conditions over time, the same set of experiments was repeated on different days). The nonclassical region of the Svetlichny inequality was subdivided into ten finer partitions; this allowed a more detailed exploration of how variations in gate composition, connectivity, and circuit depth affect the simulation of nonlocality on a quantum processor (Fig.~\ref{fig: 3-qubit_Svetlichny_Triplet_1-2-3})}.

\begin{figure*}[!h]
     \centering
     \begin{subfigure}{0.45\textwidth}
         \centering
         \includegraphics[width=\textwidth]{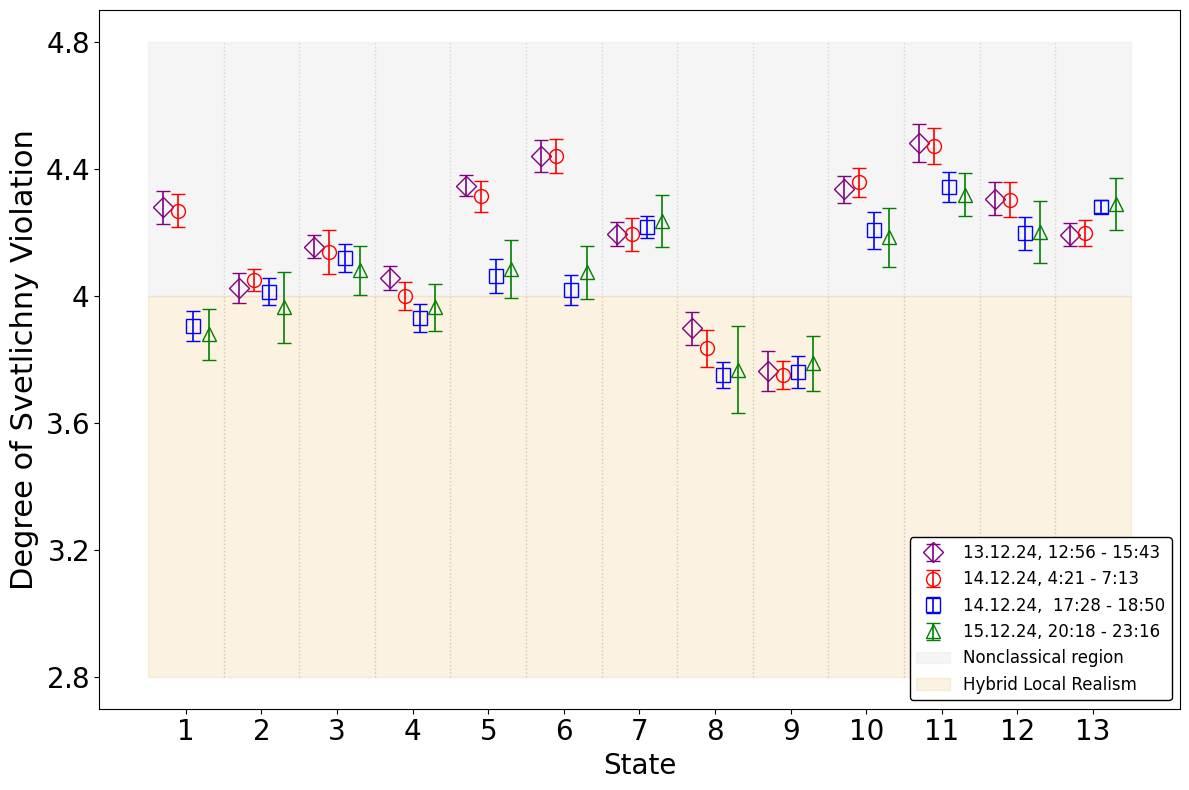}
         \caption{Triplet with qubit index 3-4-5.}
         \label{fig:3-4-5}
     \end{subfigure}
     \hfill
     \begin{subfigure}{0.45\textwidth}
         \centering
         \includegraphics[width=\textwidth]{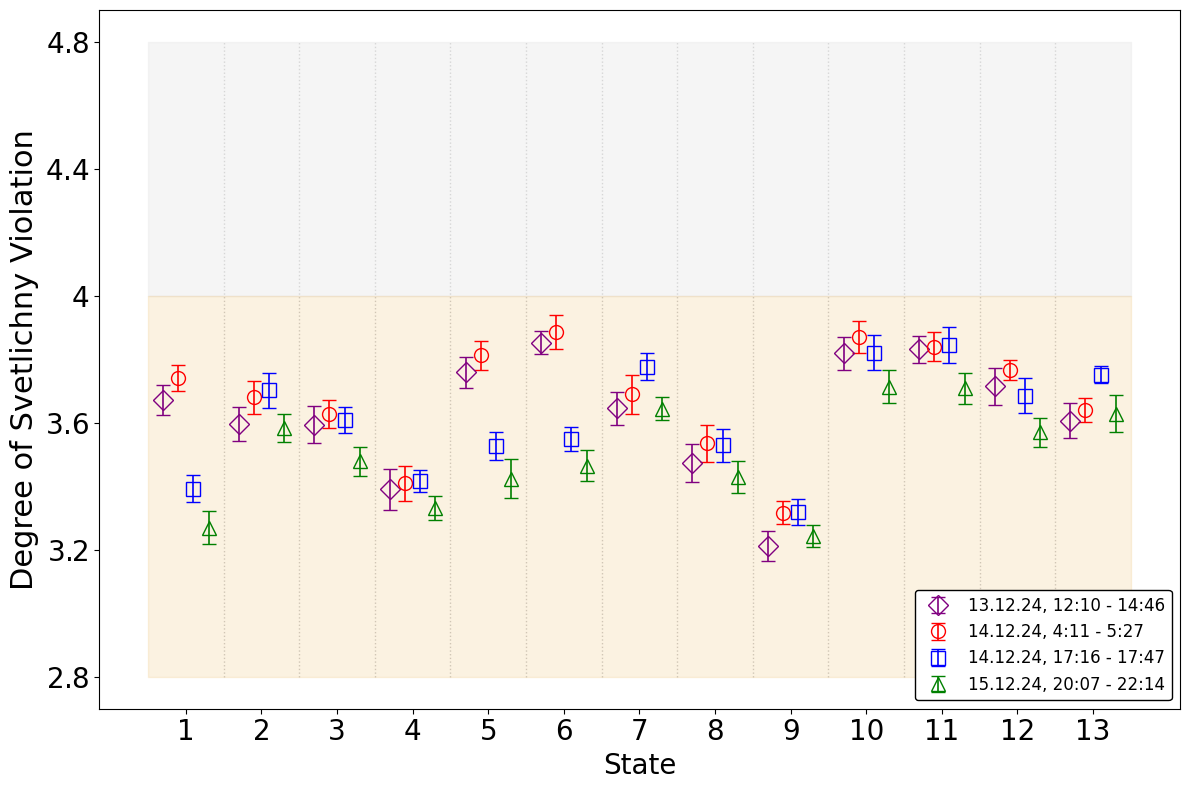}
         \caption{Triplet with qubit index 9-10-11.}
         \label{fig:9-10-11}
     \end{subfigure}
     \hfill
     \begin{subfigure}{0.45\textwidth}
        \centering
         \includegraphics[width=\textwidth]{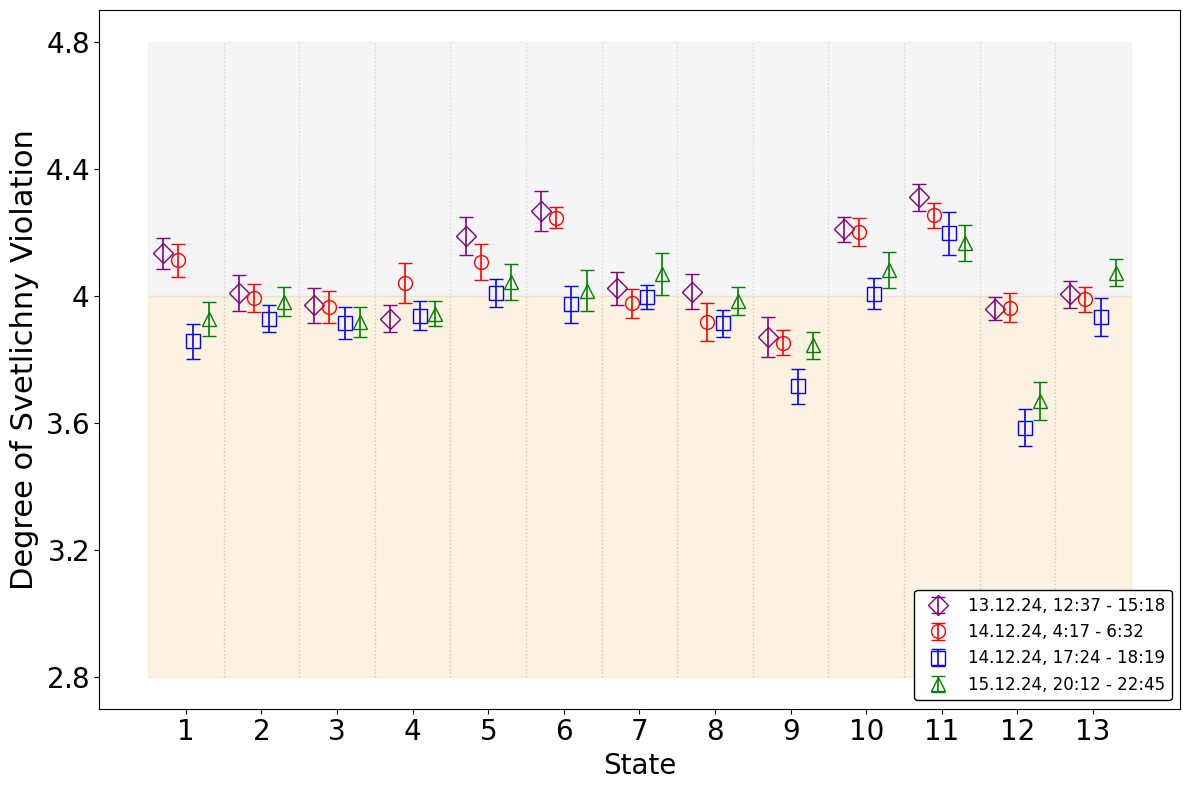}
         \caption{Triplet with qubit index 17-18-19.}
         \label{fig:17-18-19}
     \end{subfigure}
     \hfill
     \begin{subfigure}{0.45\textwidth}
         \centering
         \includegraphics[width=\textwidth]{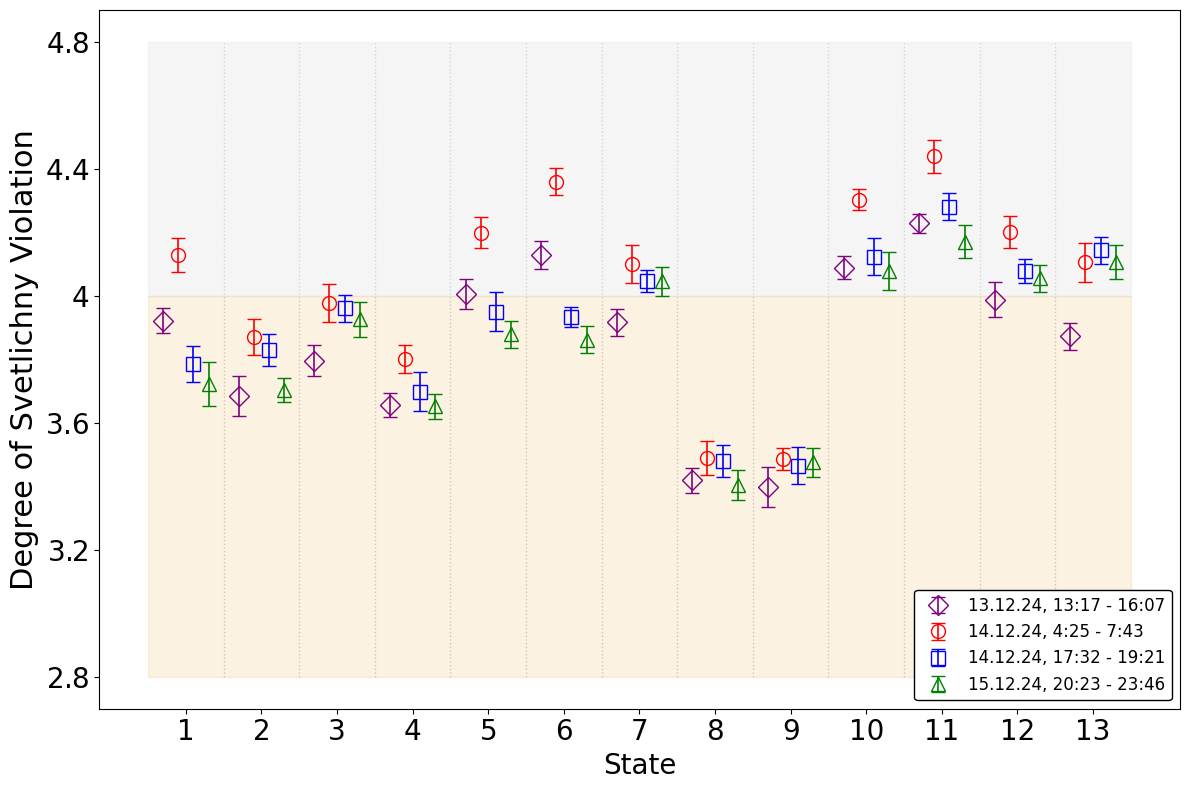}
         \caption{Triplet with qubit index 3-8-13.}
         \label{fig:3-8-13}
     \end{subfigure}
     \hfill
     \caption{Svetlichny test as a tool for evaluating the performance of different qubit triplets over time. Thirteen randomly selected states were executed on several three-qubit subsets: (a) $3$–$4$–$5$, (b) $9$–$10$–$11$, (c) $17$–$18$–$19$, and (d) $3$–$8$–$13$, at different moments in time. Qubit enumeration is shown in Fig.~\ref{fig:5-qubit_conectivity}. Marked performance differences are observed, ranging from the absence of violation in panel (b) to clear instances of violation in panel (a). In most cases, a clear time dependence is evident, as illustrated by the behavior of states $6$ and $11$ across all panels.}
    \label{fig:Comparison_Qiskit_1}       
\end{figure*}

For the comparison of the performance of the different qubit triplets, the set of quantitative evaluation measures defined in Appendix~\ref{appendix:metrics} was employed. The results indicate that all qubit triplets behave similarly to each other, within the expected experimental and statistical uncertainties (see Table~\ref{tab:triplet_deltas} and Fig.~\ref{fig: 3-qubit_Svetlichny_Triplet_1-2-3}).

\textcolor{blue}{Notice that even though the circuits are such that the qubits of the different triplets are not supposed to interact with each other, the dispersion bars in this experiment are noticeably larger than those observed when a single triplet is tested in isolation (compare Figs.~\ref{fig: 3-qubit_Svetlichny_Triplet_1-2-3} and~\ref{fig:ALL_CLIFFORD+T_IQM_AQT}). This increase is attributed to the fact that, in this experiment, all four triplets are operated simultaneously, utilizing the full register. The effective noise environment under full-load operation differs from that of isolated triplet execution, as the simultaneous manipulation of all ions in the chain may alter the collective noise landscape experienced by each individual triplet. This observation is itself relevant from a benchmarking perspective: it shows that the Svetlichny test is sensitive to operational conditions that go beyond the performance of individual qubit subsets tested in isolation.}

\newpage
\subsection{Native-gate tests across sets of qubits}
\label{ss:Native_Gates}

\textcolor{red}{In this subsection, we include experiments that shed more light on how to use tests for simulating genuine multipartite nonlocality as a benchmarking tool. The idea is now to perform Svetlichny tests on different qubit subsets of Garnet using native gates at different moments in time.}

This methodology illustrates how tests that simulate GMNL can be used to evaluate the different parts of the QPU and verify whether their performance changes over time. It is important to bear in mind that, unlike ion-trap devices, superconducting qubits are not indistinguishable. The latter involve a large number of components forming a macroscopic system whose final characteristics may depend on the fabrication process. Therefore, a substantial difference in performance between the different qubit subsets of Garnet is to be expected. Here we show how GMNL can be used to quantify this difference.

In Fig.~\ref{fig:Comparison_Qiskit_1} we randomly produced different three-qubit states that theoretically yield a maximal violation of the Svetlichny inequality Eq.~\eqref{e:SVL_N_Qubits}. This means that, should the processor be perfect, the violation value obtained should be $4\sqrt{2}$ for every test. Native gates were used, and the qubit triplet on which each test should be run was specified. For each state, each Svetlichny test was performed ten times, and this procedure was repeated on different qubit triplets and at different moments in time \textcolor{red}{(since Garnet is calibrated only once daily at a fixed hour, it is reasonable to expect that fluctuating workloads throughout the day may impact qubit fidelity)}.

\begin{figure}[h!]
\centering
\includegraphics[width=\columnwidth]{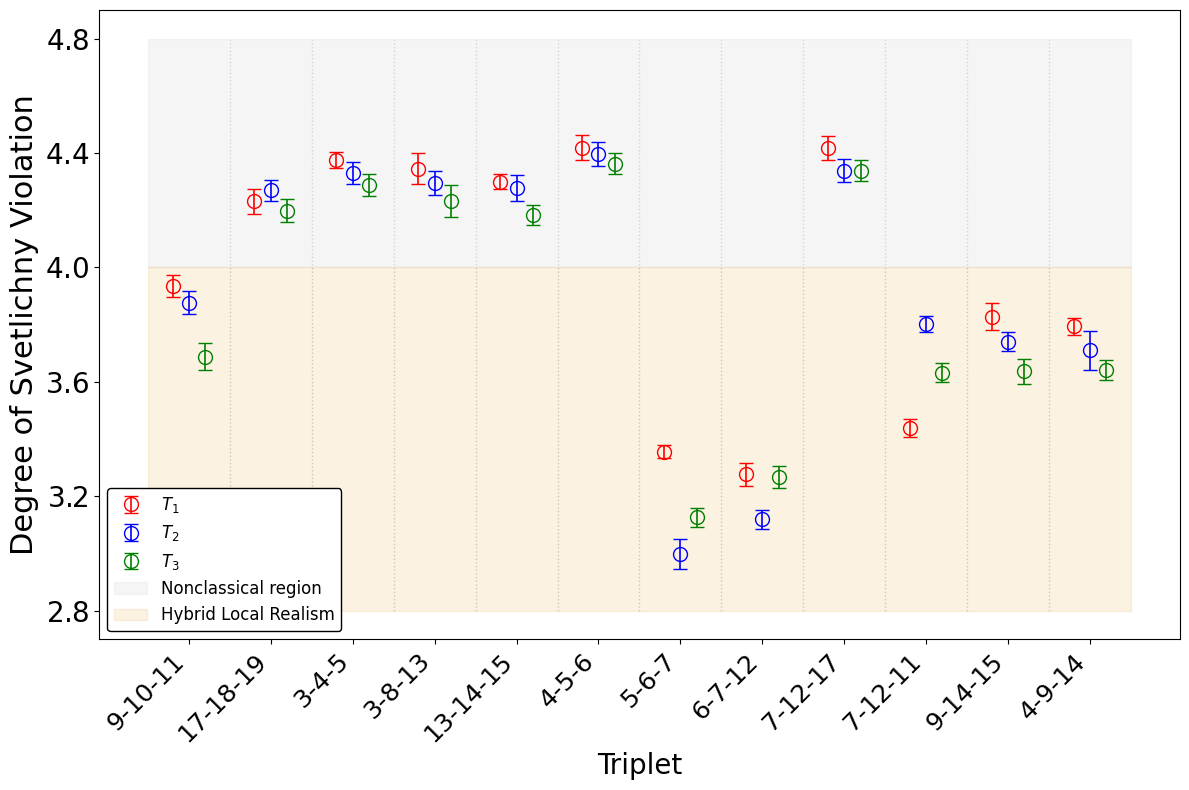}
\caption{A Svetlichny test using always the same state (LC state) is run in different triplets of qubits (indicated in the labels on the longitudinal axis) at different moments of time. Each test is repeated fifteen times. The results are displayed at different time windows $T_1= [14.12.24,(20:59-23:39)]$, $T_2=[15.12.24,(03:42-04:25)]$, and $T_3= [15.12.24,(18:10-21:15)]$. 
Our methodology allows us to detect small changes in performance, as is the case with qubits $13-14-15$ and time windows $T_{1}$ and $T_{3}$.}
\label{fig:Comparison_Qiskit_2}       
\end{figure}

The results shown in Fig.~\ref{fig:Comparison_Qiskit_1} clearly indicate that the different qubit triplets have quite different performances and that these may vary with time. The investigation of the causes of this spread is beyond the scope of this work, but it likely stems from variability in fabrication, calibration, and physical conditions among the individual qubits. Furthermore, since the GMNL test relies on the collective operation of multiple qubits, the underlying topology and the presence of crosstalk are expected to influence overall performance. The proposed methodology can therefore also be used to compare qubit subsets arranged in different topologies (see Fig.~\ref{fig:5-qubit_conectivity} for reference), a critical task for many users of QPUs based on superconducting qubits. 

The behavior observed in Fig.~\ref{fig:Comparison_Qiskit_1} is reinforced by the results displayed in Fig.~\ref{fig:Comparison_Qiskit_2}. There, a Svetlichny test for a linear cluster (LC) state (see Section~IV-B of Ref.~\cite{Cabello-Non-locality-benchmarking}) is run on different qubit triplets. Each test is repeated fifteen times with 3000 shots. The methodology is clearly sensitive to the performance of the different qubit subsets, as some of them show robust violation statistics while others do not (e.g., $5$--$6$--$7$ vs $4$--$5$--$6$). The differences between some tests indicate that, in certain cases, the methodology is also sensitive to changes over time.

\begin{figure}[!h]
    \centering
    \includegraphics[width=\linewidth]{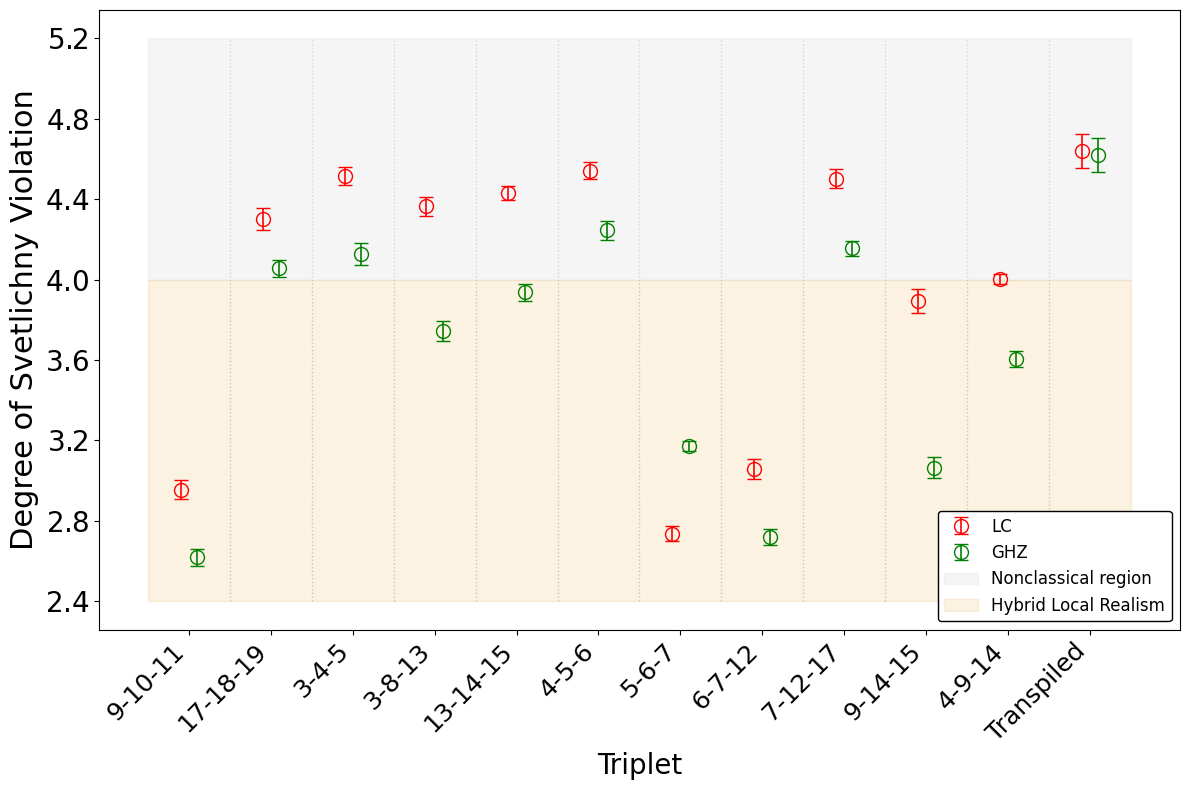}
    \caption{Evaluation of 3-qubit Svetlichny tests for LC $=[1.1.25,(18:38-19:11)]$ and GHZ $=[1.1.25,(18:17-19:14)]$ states with different triplets. For the 3-qubit case, theory predicts the same value of maximal violation of the Svetlichny inequality for both states (notice the difference in the experimental results). The state labeled ``transpiled'' on the x-axis means that the same test was run using the software optimization provided by the system. For the latter, both states gave the same violation value.}
    \label{fig:3q_GHZ_LC}
\end{figure}

In Fig.~\ref{fig:3q_GHZ_LC}, we show a comparison between the results of running Svetlichny tests using LC and GHZ states in different subsets of qubits. Theoretically, for the three-qubit case, both states reach the maximal violation value of the inequality. It is interesting that in the experiments, we have observed a difference in the violation value obtained. The experiments labeled with the qubits' triplets were performed using native gates. The one labeled with used software optimization of the system. This implies that, in the actual implementation, a different set of gates was used, and different subsets of qubits were used on each repetition of the tests. 

In Fig.~\ref{fig:3q_GHZ_LC}, we show a comparison between the results of running Svetlichny tests using LC and GHZ states on different qubit subsets. Theoretically, for the three-qubit case, both states reach the maximal violation value of the inequality. Interestingly, in the experiments we observed a difference in the violation values obtained. The experiments labeled with qubit triplets were performed using native gates. The one labeled ``transpiled'' used the software optimization provided by the system. This implies that, in the actual implementation, a different set of gates was used, and different qubit subsets were employed on each repetition of the tests.

In view of the results from the previous sections regarding Garnet, it is important to corroborate on different four-qubit subsystems (with qubit indices $3$--$4$--$5$--$6$, $6$--$7$--$12$--$17$, $13$--$14$--$18$--$19$, $8$--$9$--$14$--$13$, $14$--$15$--$19$--$18$, $3$--$4$--$9$--$8$, $6$--$7$--$12$--$11$, $4$--$9$--$14$--$18$, $8$--$13$--$14$--$15$, $5$--$10$--$15$--$19$) that the GHZ state does not exhibit Svetlichny violation in any of these cases.

\section{Conclusions}\label{s:Conclusions}

In this work, we explored nonlocality tests based on the Svetlichny inequality by executing them on quantum processors. Simulating the Svetlichny inequality on a quantum computer proves to be challenging, even for a small number of qubits. We present arguments supporting the use of this test as the basis for a benchmarking technique.

The numerical results in Section~\ref{s:Numerical} and the experimental data in Section~\ref{s:Experiments} suggest that simulating a violation of the Svetlichny inequality is significantly more challenging than violating the Mermin inequality. The experiments conducted on the Garnet processor illustrate this distinction: while Mermin violations were observed for up to five qubits, Svetlichny violations could not be achieved beyond three. These results reinforce the conclusion that Mermin inequalities are insufficient as robust benchmarks for quantifying device performance. In other words, simulating GMNL seems to be a hard task on several prototypes of NISQ devices.

Since a variant of the proposed test relies on quantum random circuits, it characterizes the capability of a given set of qubits to produce a sufficiently rich set of quantum states, using GMNL as a quantifier. This is a key difference from tests restricted to a single family of algorithms. A quantum processor might be tuned to perform well for specific state preparations or algorithms. The layer of quantum random circuits we propose makes the test unbiased with regard to any particular state or algorithm.

Our experiments on Garnet and Marmot processors indicate that the Svetlichny test can be used to assess the overall performance of a given set of qubits. Performing the test on different families of qubits within a processor can help determine which subset is expected to perform best. This is an advantage over tests based on single qubits or pairs (like those reporting one- or two-qubit gate fidelities), as the overall performance of a set cannot be reduced to the mere sum of its components due to crosstalk and other collective effects. This effect was clearly seen in Section~\ref{s:Experiments} for Marmot, given that, when using the full set of available qubits, the spread in mean values of the Svetlichny operator was considerably bigger than when testing just three or four qubits: compare the dispersion of Marmot's tests in Figs.~\ref{fig:GHZ_IQM_MERMIN_SVET} and~\ref{fig: 3-qubit_Svetlichny_Triplet_1-2-3}. In the latter, all available qubits (twelve in total) were used.

We can further enhance the test by tailoring our quantum random circuits software to incorporate a processor's native gates, as it was done in Subsection~\ref{s:Experiments}.\ref{ss:Native_Gates}. This strategic move bypasses the transpilation software layer, which often introduces an impact on performance. The use of native gates allows one to easily separate the performance limitations arising from the physical hardware from those caused by the software. In this sense, our approach can be adapted to minimize the software impact during tests targeting solely the hardware layer.

Given that the test can be performed on a given system within an hour (or less, depending on its features), it can provide insight into the evolution of qubit performance over time for calibration purposes. This was clearly seen in the results presented in Subsection~\ref{s:Experiments}.\ref{ss:Native_Gates}. In quantum processors with limited connectivity—such as those based on superconducting qubits—it can also be used to compare the performance of different topological arrangements. Thus, the proposed test is promising for assessing quantum processors across both the time and topology dimensions.

Simulating the Svetlichny inequality on quantum processors demands exponential resources. Nevertheless, the results presented here and in previous works~\cite{Granda_2025} indicate that small tests (< 10 qubits) are extremely informative about the performance of current quantum processor prototypes. Therefore, this test is a valuable tool both for benchmarking and as a figure of merit for architecture development.

The search for effective figures of merit is subtle and 
non-trivial. While one might focus on improving one- and 
two-qubit gate fidelities, advances in that metric will not 
necessarily translate to a better overall performance. This 
is why incorporating performance quantifiers that test 
subsets of several qubits can be game-changing---a philosophy 
also seen in the use of XEB (see, for 
example,~\cite{Google_XEB}). The results presented in this 
work indicate that adding genuine multipartite nonlocality 
to the battery of collective behavior tests is a step forward 
in the quest for quantum advantage. \textcolor{blue}{Moreover, the generalized Svetlichny inequality admits different classical bounds depending on the nonlocality depth $k$~\cite{Wang_2025,Baccari_2019}, making it possible to determine how many parties share genuine nonlocal correlations even when the full GMNL bound is not violated. Incorporating these intermediate bounds into the benchmarking protocol would offer a more detailed picture of processor performance beyond the binary pass/fail criterion used here}. This analysis, together with extending the tests to larger system sizes ($N > 10$), is an important direction we plan to address in future work.
\section*{Acknowledgments}

The authors gratefully acknowledge the use of the quantum systems Garnet and Marmot, operated by the Leibniz Supercomputing Centre (LRZ), for providing the computational resources for this work.

\section*{Declarations}

\subsection*{Ethics Approval}
Not applicable.

\subsection*{Funding}
Roberto Giuntini and Giuseppe Sergioli are partially supported by the projects “CORTEX – The Cost of Reasoning: Theory and Experiments” (PRIN 2022, cod. 2022ZLLR3T) and “Quantum Models for Logic, Computation and Natural Processes (Qm4Np)” (PRIN-PNRR 2022, cod. P2022A52CR), funded by the Ministry of University and Research (Italy).

Roberto Giuntini is partially funded by the T\"UV S\"UD Foundation, the Federal Ministry of Education and Research (BMBF) and the Free State of Bavaria under the Excellence Strategy of the Federal Government and the L\"ander, as well as by the Technical University of Munich-Institute for Advanced Study. 

Carla Rieger is sponsored by the Wolfgang Gentner Programme of the German Federal Ministry of Education and Research (grant no. 13E18CHA) and supported by the CERN Quantum Technology Initiative. 

The research by Luigi Iapichino is partially funded in the context of the Munich Quantum Valley, which is supported by the Bavarian state government with funds from the Hightech Agenda Bayern Plus.

The work of Tobias Valentin Bauer is partly funded from the German Federal Ministry of Research, Technology and Space via the funding program ``Quantum Technologies'' under contract number 13N16188 ``MUNIQC-SC''.

\subsection*{DATA AVAILABILITY}
\label{s:CODE_AVAILABILITY}
The data that support the findings of this article are openly available on \href{https://github.com/carlasophie/nonlocality-depolarizing-noise}{GitHub}.



\appendix


\section{Appendix: Shot-Number Stability Analysis}
\addcontentsline{toc}{section}{Review: Shot-Number Stability Analysis}
\phantomsection
\label{appendix:shots}

\begin{figure}[h!]
        \centering
        \includegraphics[width=\linewidth]{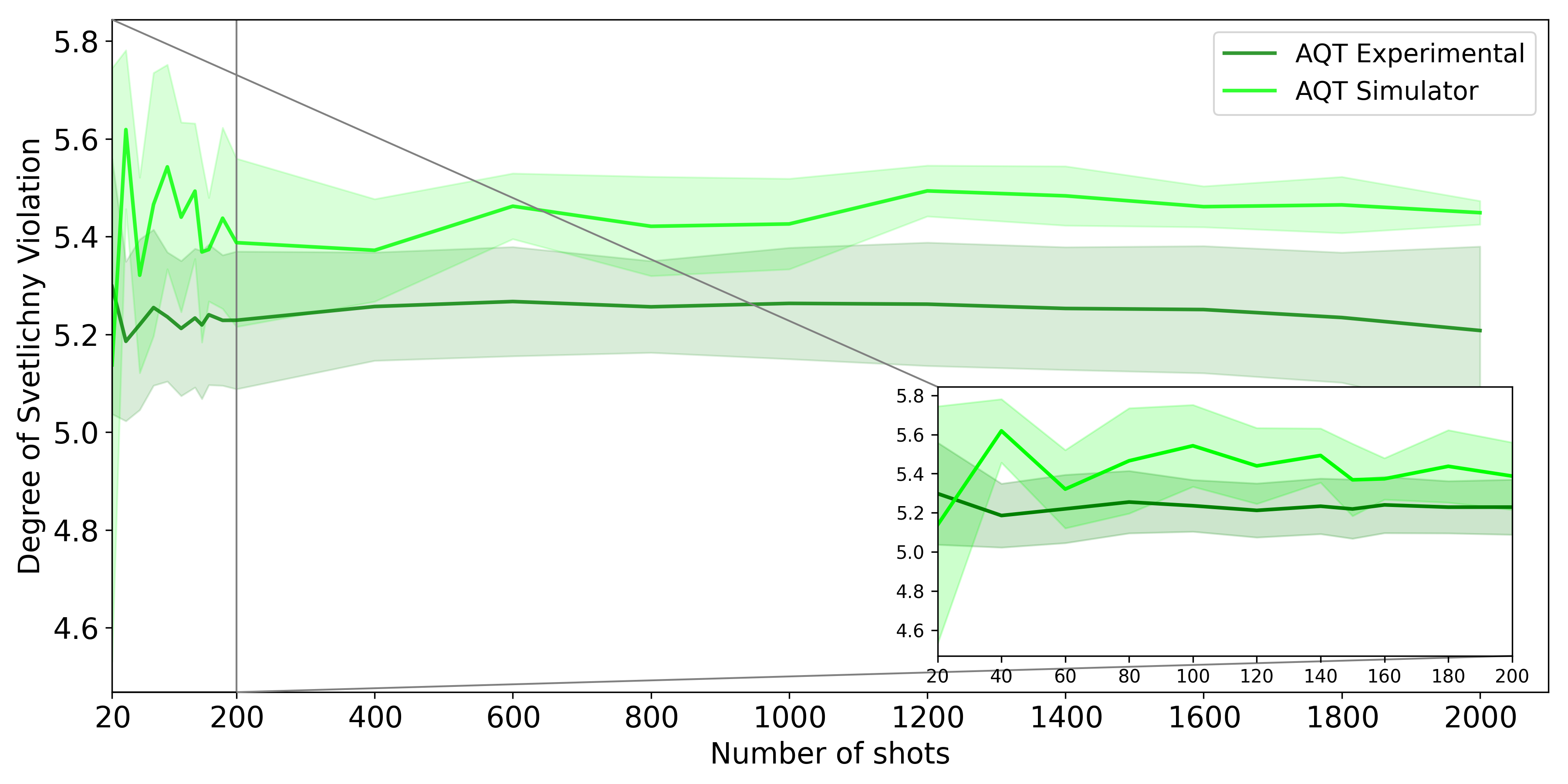}
        \caption{Svetlichny violation for a three-qubit system on Marmot, including both experimental data and simulations, as a function of the number of \emph{shots}. Shaded regions correspond to the confidence interval (mean $\pm$\,std).}
        \label{fig:Svetlichny_diferent_shots_3Q}
\end{figure}

To examine the statistical stability of Marmot under these conditions, we performed an additional test in the three-qubit case for the Svetlichny inequality, increasing the effective number of \textit{shots} from 20 to 2000 (see Fig.~\ref{fig:Svetlichny_diferent_shots_3Q}). Each shot is an independent measurement; therefore, the increase was achieved by submitting several identical jobs, each with a maximum of 200 shots. This allowed us to evaluate how increasing the number of measurements affects the variance and stability of the estimated violation values. Finally, as the number of qubits increases, larger standard deviations are expected due to device noise, and because the number of correlations required to evaluate the inequality grows. 

\section{Appendix: Evaluation methods}
\addcontentsline{toc}{section}{Review: Evaluation methods}
\phantomsection
\label{appendix:metrics}

This appendix provides the definitions of the quantitative evaluation measures employed to compare experimental results, simulations, and theoretical predictions across all devices.

Throughout this work, we denote by $S_i$ the $i$-th representative state, selected either from 
the subdivision of the violation interval (as in the Clifford+\texttt{T} analysis) or from the 
discrete set of accessible violation values (as in the Clifford group).  
Likewise, $T_j$ denotes the $j$-th triplet of qubits tested in the device.  
Thus, each pair $(S_i, T_j)$ uniquely identifies a specific experiment corresponding to a given 
state–triplet configuration.

\begin{enumerate} 

    \item \textbf{Absolute deviations.}  To quantify deviations from theoretical predictions, we define the absolute deviation
    \begin{equation}
        \delta_\mathrm{theo}(S_i,T_j)= \left|\mu_{S_i,T_j}-V_{\mathrm{theo},S_i}\right|.
        \label{eq:delta_theo}
    \end{equation}

    When several triplets are considered, the reported deviation corresponds to the average of $\delta(S_i,T_j)$ over the $K$ tested triplets.
    
    To compare the experimental outcomes obtained on different quantum architectures, the absolute difference between the mean violation values measured on Marmot and Garnet is defined by
    \begin{equation}
        \Delta(S_i,T_j)= \left|\mu_{\mathrm{AQT}}(S_i,T_j)-\mu_{\mathrm{IQM}}(S_i,T_j)\right|.
        \label{eq:delta_plat}
    \end{equation}
        
    \item \textbf{Mean relative error} quantifies the percentage deviation between the experimental value and its corresponding theoretical prediction for each representative state $S_i$:

    \begin{equation}
        \varepsilon_{\mathrm{rel}}(S_i, T_j) = \frac{|\mu_{S_i, T_j} - V_{\mathrm{theo}_ {S_i}}|}{V_{\mathrm{theo}_ {S_i}}} \times 100\% .
        \label{e:error_relativo}
    \end{equation}
    
    When several qubit triplets are tested for the same representative state (as in the triplet benchmark of Section~\ref{ss:Triplets}), the reported value corresponds to the average of $\varepsilon_{\mathrm{rel}}(S_i,T_j)$ over the $K$ tested triplets.

    \item \textbf{Normalized violation ratio} quantifies the relative performance of a given experimental configuration by normalizing the measured violation between the classical and quantum bounds:
    \begin{equation}
        R(S_i,T_j) = \frac{V_{\mathrm{exp}_{S_i,T_j}} - V_{\mathrm{LHV}}}{V_{\mathrm{ideal}_{S_i}} - V_{\mathrm{LHV}}}.
    \end{equation}

    Here, $V_{\mathrm{LHV}}$ denotes the bound imposed by local hidden-variable (LHV) theories, while $V_{\mathrm{ideal}}(S_i)$ corresponds to the maximum quantum violation value associated with the representative state $S_i$.

    While $R = 1$ indicates that the experimental value coincides with the expected violation for the corresponding representative state, negative values of $R$ indicate results below the LHV bound.
\end{enumerate}

\section{Appendix: Supplementary Tables}
\addcontentsline{toc}{section}{Tables}
\label{appendix:Tables}

\onecolumn

\begin{table*}[htbp]
\centering
\small
\caption{Theoretical, experimental, and simulated results for the Mermin and Svetlichny inequalities using Clifford+\texttt{T} states from 3 to 5 qubits.
For Garnet, both the experimental data and the \texttt{FakeAdonis} simulator outputs are reported.
For Marmot, results from the Marmot processor and its simulator (\texttt{AQT-Provider}) are included.
Mean values and standard deviations ($\mu \pm \sigma$) are computed from 20 experimental repetitions, together with the metrics $\varepsilon_{\mathrm{rel}}$ and $R$. Negative values of $R$ indicate results below the corresponding HLR bound, while entries marked with ``X'' denote cases in which $R$ is not defined because the ideal violation value coincides with the bound.}
\label{tab:CliffordT_results}

\begin{tabular}{lccccccccccc}
\toprule

& & & &
\multicolumn{2}{c}{\textbf{Garnet ($\mu\pm\sigma$)}} &
\multicolumn{2}{c}{\textbf{Marmot ($\mu\pm\sigma$)}} &
\multicolumn{2}{c}{\textbf{Garnet}} &
\multicolumn{2}{c}{\textbf{Marmot}} \\

\cmidrule(lr){5-6}
\cmidrule(lr){7-8}
\cmidrule(lr){9-10}
\cmidrule(lr){11-12}

\textbf{}
& \textbf{$N$}
& \textbf{$S$}
& \textbf{Theor.}
& \textbf{Exp.} & \textbf{Sim.}
& \textbf{Exp.} & \textbf{Sim.}
& $\boldsymbol{\varepsilon_{\mathrm{rel}}}$ & $\boldsymbol{R}$
& $\boldsymbol{\varepsilon_{\mathrm{rel}}}$ & $\boldsymbol{R}$
\\

\midrule

\multirow{18}{*}{\rotatebox{90}{Mermin}}
& \multirow{6}{*}{3}
& 1 & 1    & 0.87 $\pm$ 0.01 & 0.80 $\pm$ 0.01 & 0.88 $\pm$ 0.10 & 1.00 $\pm$ 0.01 & 12.5 & X      & 11.8 & X     \\
&& 2 & 1.2  & 0.89 $\pm$ 0.05 & 0.84 $\pm$ 0.02 & 0.99 $\pm$ 0.04 & 1.11 $\pm$ 0.01 & 26.2 & $-$0.57 & 16.9 & $-$0.01 \\
&& 3 & 1.4  & 0.99 $\pm$ 0.05 & 0.92 $\pm$ 0.02 & 1.13 $\pm$ 0.08 & 1.32 $\pm$ 0.01 & 29.4 & $-$0.03 & 18.9 & 0.33  \\
&& 4 & 1.6  & 1.23 $\pm$ 0.08 & 1.00 $\pm$ 0.02 & 1.39 $\pm$ 0.10 & 1.52 $\pm$ 0.01 & 23.4 & 0.38  & 13.1 & 0.64  \\
&& 5 & 1.8  & 1.23 $\pm$ 0.06 & 1.24 $\pm$ 0.03 & 1.56 $\pm$ 0.09 & 1.65 $\pm$ 0.01 & 31.9 & 0.28  & 13.2 & 0.70  \\
&& 6 & 2    & 1.47 $\pm$ 0.10 & 1.46 $\pm$ 0.02 & 1.81 $\pm$ 0.10 & 1.93 $\pm$ 0.01 & 26.4 & 0.47  & 9.4  & 0.81  \\

\addlinespace

& \multirow{6}{*}{4}
& 1 & 1    & 0.82 $\pm$ 0.03 & 0.73 $\pm$ 0.01 & 0.95 $\pm$ 0.06 & 1.00 $\pm$ 0.03 & 18.4 & X      & 4.3  & X     \\
&& 2 & 1.36 & 0.83 $\pm$ 0.08 & 0.92 $\pm$ 0.01 & 1.17 $\pm$ 0.11 & 1.28 $\pm$ 0.05 & 38.9 & $-$0.47 & 13.8 & 0.47  \\
&& 3 & 1.72 & 1.02 $\pm$ 0.04 & 1.07 $\pm$ 0.01 & 1.46 $\pm$ 0.09 & 1.58 $\pm$ 0.06 & 40.6 & 0.03  & 14.9 & 0.64  \\
&& 4 & 2.09 & 1.19 $\pm$ 0.05 & 1.28 $\pm$ 0.02 & 1.69 $\pm$ 0.09 & 1.89 $\pm$ 0.06 & 43.0 & 0.17  & 18.8 & 0.63  \\
&& 5 & 2.44 & 1.34 $\pm$ 0.08 & 1.34 $\pm$ 0.01 & 2.13 $\pm$ 0.10 & 2.27 $\pm$ 0.05 & 44.9 & 0.24  & 12.5 & 0.78  \\
&& 6 & 2.82 & 1.86 $\pm$ 0.05 & 1.88 $\pm$ 0.01 & 2.35 $\pm$ 0.12 & 2.64 $\pm$ 0.06 & 34.2 & 0.47  & 16.9 & 0.73  \\

\addlinespace

& \multirow{6}{*}{5}
& 1 & 1    & 0.66 $\pm$ 0.05 & 0.59 $\pm$ 0.01 & 0.94 $\pm$ 0.06 & 0.96 $\pm$ 0.01 & 33.5 & X      & 5.7  & X     \\
&& 2 & 1.6  & 0.92 $\pm$ 0.05 & 0.98 $\pm$ 0.01 & 1.21 $\pm$ 0.09 & 1.43 $\pm$ 0.01 & 42.2 & $-$0.13 & 24.3 & 0.35  \\
&& 3 & 2.2  & 0.99 $\pm$ 0.04 & 1.14 $\pm$ 0.01 & 1.74 $\pm$ 0.09 & 1.97 $\pm$ 0.01 & 54.7 & 0     & 20.7 & 0.61  \\
&& 4 & 2.8  & 1.27 $\pm$ 0.06 & 1.51 $\pm$ 0.01 & 2.19 $\pm$ 0.14 & 2.43 $\pm$ 0.01 & 54.6 & 0.15  & 21.7 & 0.66  \\
&& 5 & 3.4  & 1.43 $\pm$ 0.06 & 1.61 $\pm$ 0.01 & 2.31 $\pm$ 0.18 & 2.63 $\pm$ 0.02 & 57.9 & 0.18  & 31.7 & 0.54  \\
&& 6 & 4    & 2.09 $\pm$ 0.09 & 2.44 $\pm$ 0.01 & 3.34 $\pm$ 0.14 & 3.63 $\pm$ 0.01 & 47.7 & 0.36  & 16.2 & 0.78  \\

\addlinespace
\midrule

\multirow{18}{*}{\rotatebox{90}{Svetlichny}}
& \multirow{6}{*}{3}
& 1 & 4    & 2.90 $\pm$ 0.24 & 3.08 $\pm$ 0.04 & 3.59 $\pm$ 0.22 & 3.92 $\pm$ 0.01 & 27.4 & X      & 10.1 & X     \\
&& 2 & 4.3  & 3.36 $\pm$ 0.16 & 2.93 $\pm$ 0.05 & 3.92 $\pm$ 0.22 & 4.09 $\pm$ 0.04 & 22.3 & $-$1.93 & 8.6  & $-$0.24 \\
&& 3 & 4.6  & 3.27 $\pm$ 0.16 & 3.26 $\pm$ 0.04 & 4.08 $\pm$ 0.19 & 4.35 $\pm$ 0.05 & 29.9 & $-$1.11 & 11.2 & $-$0.13 \\
&& 4 & 4.9  & 3.49 $\pm$ 0.12 & 3.27 $\pm$ 0.04 & 4.62 $\pm$ 0.16 & 4.70 $\pm$ 0.04 & 30.1 & $-$0.52 & 5.6  & 0.69  \\
&& 5 & 5.3  & 3.85 $\pm$ 0.10 & 3.54 $\pm$ 0.05 & 4.82 $\pm$ 0.12 & 4.96 $\pm$ 0.04 & 27.7 & $-$0.11 & 8.9  & 0.63  \\
&& 6 & 5.6  & 4.25 $\pm$ 0.13 & 4.00 $\pm$ 0.05 & 5.11 $\pm$ 0.19 & 5.36 $\pm$ 0.06 & 24.9 & 0.15  & 9.4  & 0.67  \\

\addlinespace

& \multirow{6}{*}{4}
& 1 & 8     & 4.60 $\pm$ 0.17 & 4.50 $\pm$ 0.07 & 6.87 $\pm$ 0.41 & 7.43 $\pm$ 0.09 & 42.5 & X      & 14.1 & X     \\
&& 2 & 8.6   & 5.09 $\pm$ 0.26 & 4.91 $\pm$ 0.07 & 6.92 $\pm$ 0.31 & 7.73 $\pm$ 0.08 & 41.2 & $-$4.41 & 20.1 & $-$1.64 \\
&& 3 & 9.3   & 5.70 $\pm$ 0.13 & 5.60 $\pm$ 0.07 & 7.63 $\pm$ 0.35 & 8.56 $\pm$ 0.05 & 38.8 & $-$1.74 & 18.1 & $-$0.27 \\
&& 4 & 9.9   & 5.61 $\pm$ 0.21 & 6.22 $\pm$ 0.06 & 8.17 $\pm$ 0.33 & 9.10 $\pm$ 0.06 & 43.8 & $-$1.21 & 18.1 & 0.08  \\
&& 5 & 10.6  & 6.75 $\pm$ 0.15 & 6.85 $\pm$ 0.05 & 8.48 $\pm$ 0.28 & 9.34 $\pm$ 0.04 & 36.6 & $-$0.47 & 20.3 & 0.18  \\
&& 6 & 11.3  & 7.37 $\pm$ 0.14 & 7.27 $\pm$ 0.05 & 9.27 $\pm$ 0.32 & 10.43 $\pm$ 0.06 & 34.8 & $-$0.19 & 18.0 & 0.38  \\

\addlinespace

& \multirow{6}{*}{5}
& 1 & 16    & 9.19 $\pm$ 0.42  & 8.98 $\pm$ 0.07  & 12.16 $\pm$ 0.37 & 13.25 $\pm$ 0.09 & 42.6 & X      & 24.0 & X     \\
&& 2 & 17.32 & 10.00 $\pm$ 0.23 & 9.38 $\pm$ 0.08  & 13.08 $\pm$ 0.67 & 14.69 $\pm$ 0.08 & 42.3 & $-$4.55 & 24.4 & $-$2.19 \\
&& 3 & 18.65 & 9.56 $\pm$ 0.46  & 9.47 $\pm$ 0.10  & 13.98 $\pm$ 0.40 & 16.12 $\pm$ 0.09 & 48.7 & $-$2.43 & 25.1 & $-$0.75 \\
&& 4 & 19.97 & 10.30 $\pm$ 0.19 & 11.00 $\pm$ 0.10 & 15.63 $\pm$ 0.40 & 17.77 $\pm$ 0.11 & 48.6 & $-$1.45 & 21.7 & $-$0.09 \\
&& 5 & 21.30 & 11.50 $\pm$ 0.56 & 11.30 $\pm$ 0.10 & 16.73 $\pm$ 0.42 & 18.11 $\pm$ 0.09 & 46.2 & $-$0.86 & 21.4 & 0.13  \\
&& 6 & 22.62 & 11.70 $\pm$ 0.50 & 13.40 $\pm$ 0.07 & 18.90 $\pm$ 0.49 & 20.81 $\pm$ 0.07 & 48.3 & $-$0.65 & 16.4 & 0.43  \\

\bottomrule
\end{tabular}

\end{table*}


\begin{table*}[htbp]
\centering
\small
\caption{Theoretical, experimental, and simulated results for the Mermin and Svetlichny inequalities using Clifford states from 3 to 5 qubits.
For Garnet, both the experimental data and the \texttt{FakeAdonis} simulator outputs are reported.
For Marmot, results from the Marmot processor and its simulator (\texttt{AQT-Provider}) are included.
Mean values and standard deviations ($\mu \pm \sigma$) are computed from 20 experimental repetitions, together with the metrics $\varepsilon_{\mathrm{rel}}$ and $R$. Negative values of $R$ indicate results below the corresponding HLR bound, while entries marked with ``X'' denote cases in which $R$ is not defined because the ideal violation value coincides with the bound.}
\label{tab:Clifford_results}

\begin{tabular}{lccccccccccc}
\toprule

& & & &
\multicolumn{2}{c}{\textbf{Garnet ($\mu\pm\sigma$)}} &
\multicolumn{2}{c}{\textbf{Marmot ($\mu\pm\sigma$)}} &
\multicolumn{2}{c}{\textbf{Garnet}} &
\multicolumn{2}{c}{\textbf{Marmot}} \\

\cmidrule(lr){5-6}
\cmidrule(lr){7-8}
\cmidrule(lr){9-10}
\cmidrule(lr){11-12}

\textbf{}
& \textbf{$N$}
& \textbf{$S$}
& \textbf{Theor.}
& \textbf{Exp.} & \textbf{Sim.}
& \textbf{Exp.} & \textbf{Sim.}
& $\boldsymbol{\varepsilon_{\mathrm{rel}}}$ & $\boldsymbol{R}$
& $\boldsymbol{\varepsilon_{\mathrm{rel}}}$ & $\boldsymbol{R}$
\\

\midrule

\multirow{12}{*}{\rotatebox{90}{Mermin}} & \multirow{3}{*}{3}
& 1 & 1     & 0.86 $\pm$ 0.01 & 0.80 $\pm$ 0.01 & 0.99 $\pm$ 0.04 & 0.98 $\pm$ 0.01 & 14.4 & X     & 0.3  & X    \\
&& 2 & 1.41  & 1.05 $\pm$ 0.03 & 1.04 $\pm$ 0.01 & 1.32 $\pm$ 0.09 & 1.30 $\pm$ 0.01 & 25.8 & 0.12  & 6.5  & 0.78 \\
&& 3 & 2     & 1.48 $\pm$ 0.05 & 1.47 $\pm$ 0.01 & 1.82 $\pm$ 0.07 & 1.89 $\pm$ 0.01 & 25.9 & 0.48  & 8.6  & 0.83 \\

\addlinespace

& \multirow{4}{*}{4}
& 1 & 1     & 0.74 $\pm$ 0.04 & 0.63 $\pm$ 0.04 & 1.00 $\pm$ 0.04 & 0.99 $\pm$ 0.01 & 25.5 & X      & 0.4  & X    \\
&& 2 & 1.41  & 0.98 $\pm$ 0.05 & 0.84 $\pm$ 0.02 & 1.30 $\pm$ 0.11 & 1.37 $\pm$ 0.01 & 30.5 & $-$0.04 & 7.7  & 0.73 \\
&& 3 & 2     & 1.45 $\pm$ 0.06 & 1.29 $\pm$ 0.01 & 1.89 $\pm$ 0.05 & 1.92 $\pm$ 0.01 & 27.4 & 0.45  & 5.3  & 0.89 \\
&& 4 & 2.82  & 1.79 $\pm$ 0.05 & 1.80 $\pm$ 0.02 & 2.49 $\pm$ 0.10 & 2.68 $\pm$ 0.01 & 36.7 & 0.43  & 11.6 & 0.82 \\

\addlinespace

& \multirow{5}{*}{5}
& 1 & 1     & 0.74 $\pm$ 0.02 & 0.74 $\pm$ 0.01 & 0.84 $\pm$ 0.16 & 0.99 $\pm$ 0.01 & 26.2 & X     & 16.3 & X    \\
&& 2 & 1.41  & 1.13 $\pm$ 0.05 & 1.05 $\pm$ 0.02 & 1.24 $\pm$ 0.07 & 1.36 $\pm$ 0.01 & 20.2 & 0.31  & 12.4 & 0.57 \\
&& 3 & 2     & 1.48 $\pm$ 0.08 & 1.46 $\pm$ 0.01 & 1.66 $\pm$ 0.09 & 1.90 $\pm$ 0.01 & 26.1 & 0.48  & 16.8 & 0.66 \\
&& 4 & 2.82  & 1.70 $\pm$ 0.04 & 1.80 $\pm$ 0.01 & 2.37 $\pm$ 0.10 & 2.68 $\pm$ 0.01 & 39.7 & 0.39  & 16.0 & 0.75 \\
&& 5 & 4     & 2.07 $\pm$ 0.09 & 2.43 $\pm$ 0.01 & 3.40 $\pm$ 0.16 & 3.73 $\pm$ 0.01 & 48.2 & 0.36  & 15.0 & 0.79 \\

\addlinespace
\midrule

\multirow{10}{*}{\rotatebox{90}{Svetlichny}}
& \multirow{2}{*}{3}
& 1 & 4     & 3.27 $\pm$ 0.03 & 3.24 $\pm$ 0.03 & 3.65 $\pm$ 0.29 & 3.87 $\pm$ 0.01 & 18.2 & X     & 8.62 & X    \\
&& 2 & 5.65  & 4.09 $\pm$ 0.15 & 4.16 $\pm$ 0.04 & 5.28 $\pm$ 0.10 & 5.27 $\pm$ 0.04 & 27.8 & 0.05  & 6.49 & 0.78 \\

\addlinespace

& \multirow{4}{*}{4}
& 1 & 4     & 2.80 $\pm$ 0.07 & 2.81 $\pm$ 0.04 & 3.86 $\pm$ 0.10 & 3.88 $\pm$ 0.05 & 29.9 & $-$1.30 & 3.6  & $-$1.03 \\
&& 2 & 5.65  & 3.23 $\pm$ 0.06 & 3.98 $\pm$ 0.06 & 4.87 $\pm$ 0.45 & 5.36 $\pm$ 0.06 & 42.9 & $-$2.03 & 13.9 & $-$1.33 \\
&& 3 & 8     & 4.66 $\pm$ 0.17 & 4.74 $\pm$ 0.06 & 7.42 $\pm$ 0.31 & 7.72 $\pm$ 0.05 & 41.7 & X      & 7.1  & X    \\
&& 4 & 11.31 & 7.07 $\pm$ 0.20 & 6.23 $\pm$ 0.07 & 9.81 $\pm$ 0.52 & 10.63 $\pm$ 0.07 & 37.5 & $-$0.28 & 13.3 & 0.55 \\

\addlinespace

& \multirow{4}{*}{5}
& 1 & 8     & 5.78 $\pm$ 0.21 & 4.58 $\pm$ 0.09 & 7.61 $\pm$ 0.61  & 7.63 $\pm$ 0.04  & 27.8 & $-$1.28 & 4.8  & $-$1.04 \\
&& 2 & 11.31 & 6.25 $\pm$ 0.16 & 6.17 $\pm$ 0.08 & 8.15 $\pm$ 0.83  & 10.95 $\pm$ 0.09 & 44.8 & $-$2.08 & 28.0 & $-$1.68 \\
&& 3 & 16    & 7.55 $\pm$ 0.27 & 8.30 $\pm$ 0.09 & 11.25 $\pm$ 0.84 & 14.39 $\pm$ 0.09 & 52.8 & X      & 29.7 & X    \\
&& 4 & 22.62 & 10.40 $\pm$ 0.46 & 11.30 $\pm$ 0.09 & 17.56 $\pm$ 0.68 & 20.88 $\pm$ 0.08 & 54.1 & $-$0.85 & 22.4 & 0.23 \\

\bottomrule
\end{tabular}

\end{table*}



\begin{table*}[htbp]
\centering
\small
\caption{
Mean experimental values ($\mu \pm \sigma$) of the Svetlichny inequality and corresponding absolute deviations ($\delta_\mathrm{theo}$) from the ideal theoretical values for each state--triplet configuration $(S_i, T_j)$ in the Clifford+\texttt{T} analysis, obtained from 20 repeated experimental executions.
}

\begin{tabular}{cccccccccc}

\toprule

&
&
\multicolumn{4}{c}{\textbf{Marmot Experimental ($\mu \pm \sigma$)}}
&
\multicolumn{4}{c}{$\boldsymbol{\delta_\mathrm{theo}(S_i,T_j)}$}
\\

\cmidrule(lr){3-6}
\cmidrule(lr){7-10}

\textbf{State}
&
\textbf{\begin{tabular}[c]{@{}c@{}}Theor.\\ Value\end{tabular}}
&
$\mathbf{T_1}$
&
$\mathbf{T_2}$
&
$\mathbf{T_3}$
&
$\mathbf{T_4}$
&
$\mathbf{T_1}$
&
$\mathbf{T_2}$
&
$\mathbf{T_3}$
&
$\mathbf{T_4}$
\\

\midrule

1 & 4
& $3.6 \pm 0.42$
& $3.4 \pm 0.49$
& $3.5 \pm 0.35$
& $3.6 \pm 0.65$
& 0.34
& 0.59
& 0.51
& 0.36
\\

2 & 4.2
& $3.6 \pm 0.53$
& $3.5 \pm 0.49$
& $3.6 \pm 0.40$
& $3.7 \pm 0.28$
& 0.60
& 0.65
& 0.62
& 0.51
\\

3 & 4.4
& $3.7 \pm 0.58$
& $3.4 \pm 0.42$
& $3.7 \pm 0.41$
& $3.7 \pm 0.37$
& 0.72
& 1.0
& 0.71
& 0.69
\\

4 & 4.6
& $3.9 \pm 0.53$
& $3.9 \pm 0.42$
& $3.5 \pm 0.50$
& $3.5 \pm 0.68$
& 0.67
& 0.72
& 1.1
& 1.1
\\

5 & 4.8
& $4.0 \pm 0.39$
& $3.9 \pm 0.31$
& $3.8 \pm 0.29$
& $3.8 \pm 0.36$
& 0.83
& 0.91
& 0.97
& 1.0
\\

6 & 5.0
& $4.5 \pm 0.45$
& $4.4 \pm 0.50$
& $4.4 \pm 0.46$
& $4.1 \pm 0.45$
& 0.49
& 0.58
& 0.63
& 0.86
\\

7 & 5.2
& $4.6 \pm 0.30$
& $4.5 \pm 0.30$
& $4.6 \pm 0.30$
& $4.4 \pm 0.51$
& 0.57
& 0.70
& 0.61
& 0.77
\\

8 & 5.4
& $4.4 \pm 0.85$
& $4.2 \pm 0.79$
& $4.2 \pm 0.84$
& $4.2 \pm 0.79$
& 1.0
& 1.2
& 1.2
& 1.2
\\

9 & 5.6
& $4.8 \pm 0.36$
& $4.9 \pm 0.34$
& $5.0 \pm 0.38$
& $4.5 \pm 0.95$
& 0.76
& 0.74
& 0.60
& 1.1
\\

10 & 5.65
& $4.8 \pm 0.48$
& $4.6 \pm 0.38$
& $4.5 \pm 0.38$
& $4.4 \pm 0.50$
& 0.88
& 1.0
& 1.2
& 1.3
\\

\bottomrule

\end{tabular}

\label{tab:triplet_deltas}

\end{table*}

\clearpage
\newpage
\twocolumn
\small
\bibliography{mybibliography}

\begin{thebibliography}{10}
\urlstyle{rm}
\expandafter\ifx\csname url\endcsname\relax
  \def\url#1{\texttt{#1}}\fi
\expandafter\ifx\csname urlprefix\endcsname\relax\def\urlprefix{URL }\fi
\expandafter\ifx\csname doiprefix\endcsname\relax\def\doiprefix{DOI: }\fi
\providecommand{\bibinfo}[2]{#2}
\providecommand{\eprint}[2][]{\url{#2}}

\bibitem{Deutsch1992}
\bibinfo{author}{Deutsch, D.} \& \bibinfo{author}{Jozsa, R.}
\newblock \bibinfo{journal}{\bibinfo{title}{Rapid solution of problems by
  quantum computation}}.
\newblock {\emph{\JournalTitle{Proceedings of the Royal Society of London.
  Series A: Mathematical and Physical Sciences}}}
  \textbf{\bibinfo{volume}{439}}, \bibinfo{pages}{553--558},
  \doiprefix\url{10.1098/rspa.1992.0167} (\bibinfo{year}{1992}).

\bibitem{Shor1997}
\bibinfo{author}{Shor, P.~W.}
\newblock \bibinfo{journal}{\bibinfo{title}{Polynomial-time algorithms for
  prime factorization and discrete logarithms on a quantum computer}}.
\newblock {\emph{\JournalTitle{SIAM Journal on Computing}}}
  \textbf{\bibinfo{volume}{26}}, \bibinfo{pages}{1484--1509},
  \doiprefix\url{10.1137/S0097539795293172} (\bibinfo{year}{1997}).

\bibitem{Simon1997}
\bibinfo{author}{Simon, D.~R.}
\newblock \bibinfo{journal}{\bibinfo{title}{On the power of quantum
  computation}}.
\newblock {\emph{\JournalTitle{SIAM Journal on Computing}}}
  \textbf{\bibinfo{volume}{26}}, \bibinfo{pages}{1474--1483},
  \doiprefix\url{10.1137/S0097539796298637} (\bibinfo{year}{1997}).

\bibitem{Riste2017}
\bibinfo{author}{Rist\`{e}, D.} \emph{et~al.}
\newblock \bibinfo{journal}{\bibinfo{title}{Demonstration of quantum advantage
  in machine learning}}.
\newblock {\emph{\JournalTitle{npj Quantum Information}}}
  \textbf{\bibinfo{volume}{3}}, \bibinfo{pages}{16},
  \doiprefix\url{10.1038/s41534-017-0017-3} (\bibinfo{year}{2017}).

\bibitem{Feynman1982}
\bibinfo{author}{Feynman, R.~P.}
\newblock \bibinfo{journal}{\bibinfo{title}{Simulating physics with
  computers}}.
\newblock {\emph{\JournalTitle{International Journal of Theoretical Physics}}}
  \textbf{\bibinfo{volume}{21}}, \bibinfo{pages}{467--488},
  \doiprefix\url{10.1007/BF02650179} (\bibinfo{year}{1982}).

\bibitem{Lloyd1996}
\bibinfo{author}{Lloyd, S.}
\newblock \bibinfo{journal}{\bibinfo{title}{Universal quantum simulators}}.
\newblock {\emph{\JournalTitle{Science}}} \textbf{\bibinfo{volume}{273}},
  \bibinfo{pages}{1073--1078}, \doiprefix\url{10.1126/science.273.5278.1073}
  (\bibinfo{year}{1996}).

\bibitem{Bernien2017}
\bibinfo{author}{Bernien, H.} \emph{et~al.}
\newblock \bibinfo{journal}{\bibinfo{title}{Probing many-body dynamics on a
  51-atom quantum simulator}}.
\newblock {\emph{\JournalTitle{Nature}}} \textbf{\bibinfo{volume}{551}},
  \bibinfo{pages}{579--584}, \doiprefix\url{10.1038/nature24622}
  (\bibinfo{year}{2017}).

\bibitem{Cubitt2018}
\bibinfo{author}{Cubitt, T.~S.}, \bibinfo{author}{Montanaro, A.} \&
  \bibinfo{author}{Piddock, S.}
\newblock \bibinfo{journal}{\bibinfo{title}{Universal quantum {H}amiltonians}}.
\newblock {\emph{\JournalTitle{Proceedings of the National Academy of
  Sciences}}} \textbf{\bibinfo{volume}{115}}, \bibinfo{pages}{9497--9502},
  \doiprefix\url{10.1073/pnas.1804949115} (\bibinfo{year}{2018}).

\bibitem{Bloch2012}
\bibinfo{author}{Bloch, I.}, \bibinfo{author}{Dalibard, J.} \&
  \bibinfo{author}{Nascimb\`{e}ne, S.}
\newblock \bibinfo{journal}{\bibinfo{title}{Quantum simulations with ultracold
  quantum gases}}.
\newblock {\emph{\JournalTitle{Nature Physics}}} \textbf{\bibinfo{volume}{8}},
  \bibinfo{pages}{267--276}, \doiprefix\url{10.1038/nphys2259}
  (\bibinfo{year}{2012}).

\bibitem{Abanin2025}
\bibinfo{author}{Abanin, D.~A.}, \bibinfo{author}{AI, G.~Q.} \&
  \bibinfo{author}{{Collaborators}}.
\newblock \bibinfo{journal}{\bibinfo{title}{Observation of constructive
  interference at the edge of quantum ergodicity}}.
\newblock {\emph{\JournalTitle{Nature}}} \textbf{\bibinfo{volume}{646}},
  \bibinfo{pages}{825--830}, \doiprefix\url{10.1038/s41586-025-09526-6}
  (\bibinfo{year}{2025}).

\bibitem{Tura2014}
\bibinfo{author}{Tura, J.} \emph{et~al.}
\newblock \bibinfo{journal}{\bibinfo{title}{Detecting nonlocality in many-body
  quantum states}}.
\newblock {\emph{\JournalTitle{Science}}} \textbf{\bibinfo{volume}{344}},
  \bibinfo{pages}{1256--1258}, \doiprefix\url{10.1126/science.1247715}
  (\bibinfo{year}{2014}).

\bibitem{Mermin_1990}
\bibinfo{author}{Mermin, N.~D.}
\newblock \bibinfo{journal}{\bibinfo{title}{Extreme quantum entanglement in a
  superposition of macroscopically distinct states}}.
\newblock {\emph{\JournalTitle{Phys. Rev. Lett.}}}
  \textbf{\bibinfo{volume}{65}}, \bibinfo{pages}{1838--1840},
  \doiprefix\url{10.1103/PhysRevLett.65.1838} (\bibinfo{year}{1990}).

\bibitem{Svetlichny_PRD}
\bibinfo{author}{Svetlichny, G.}
\newblock \bibinfo{journal}{\bibinfo{title}{Distinguishing three-body from
  two-body nonseparability by a {B}ell-type inequality}}.
\newblock {\emph{\JournalTitle{Phys. Rev. D}}} \textbf{\bibinfo{volume}{35}},
  \bibinfo{pages}{3066--3069}, \doiprefix\url{10.1103/PhysRevD.35.3066}
  (\bibinfo{year}{1987}).

\bibitem{Seevinck_Svetlichny_PRL}
\bibinfo{author}{Seevinck, M.} \& \bibinfo{author}{Svetlichny, G.}
\newblock \bibinfo{journal}{\bibinfo{title}{Bell-type inequalities for partial
  separability in $n$-particle systems and quantum mechanical violations}}.
\newblock {\emph{\JournalTitle{Phys. Rev. Lett.}}}
  \textbf{\bibinfo{volume}{89}}, \bibinfo{pages}{060401},
  \doiprefix\url{10.1103/PhysRevLett.89.060401} (\bibinfo{year}{2002}).

\bibitem{Cabello-Non-locality-benchmarking}
\bibinfo{author}{B\"onsel, J.~L.}, \bibinfo{author}{G\"uhne, O.} \&
  \bibinfo{author}{Cabello, A.}
\newblock \bibinfo{journal}{\bibinfo{title}{Generating multipartite nonlocality
  to benchmark quantum computers}}.
\newblock {\emph{\JournalTitle{Phys. Rev. A}}} \textbf{\bibinfo{volume}{111}},
  \bibinfo{pages}{012207}, \doiprefix\url{10.1103/PhysRevA.111.012207}
  (\bibinfo{year}{2025}).

\bibitem{Lanyon2014}
\bibinfo{author}{Lanyon, B.~P.} \emph{et~al.}
\newblock \bibinfo{journal}{\bibinfo{title}{Experimental violation of
  multipartite bell inequalities with trapped ions}}.
\newblock {\emph{\JournalTitle{Phys. Rev. Lett.}}}
  \textbf{\bibinfo{volume}{112}}, \bibinfo{pages}{100403},
  \doiprefix\url{10.1103/PhysRevLett.112.100403} (\bibinfo{year}{2014}).

\bibitem{Baumer2021}
\bibinfo{author}{B{\"a}umer, E.}, \bibinfo{author}{Gisin, N.} \&
  \bibinfo{author}{Tavakoli, A.}
\newblock \bibinfo{journal}{\bibinfo{title}{Demonstrating the power of quantum
  computers, certification of highly entangled measurements and scalable
  quantum nonlocality}}.
\newblock {\emph{\JournalTitle{npj Quantum Information}}}
  \textbf{\bibinfo{volume}{7}}, \bibinfo{pages}{117},
  \doiprefix\url{10.1038/s41534-021-00450-x} (\bibinfo{year}{2021}).

\bibitem{Wang_2025}
\bibinfo{author}{Wang, K.} \emph{et~al.}
\newblock \bibinfo{journal}{\bibinfo{title}{Probing many-body bell correlation
  depth with superconducting qubits}}.
\newblock {\emph{\JournalTitle{Phys. Rev. X}}} \textbf{\bibinfo{volume}{15}},
  \bibinfo{pages}{021024}, \doiprefix\url{10.1103/PhysRevX.15.021024}
  (\bibinfo{year}{2025}).

\bibitem{Granda_2025}
\bibinfo{author}{Granda~Arango, A.~C.}, \bibinfo{author}{Holik, F.~H.},
  \bibinfo{author}{Giuntini, R.}, \bibinfo{author}{Freytes, H.} \&
  \bibinfo{author}{Sergioli, G.}
\newblock \bibinfo{journal}{\bibinfo{title}{Distribution of nonlocality on
  quantum random circuits}}.
\newblock {\emph{\JournalTitle{Phys. Rev. A}}} \textbf{\bibinfo{volume}{112}},
  \bibinfo{pages}{062427}, \doiprefix\url{10.1103/8nbw-bvj2}
  (\bibinfo{year}{2025}).

\bibitem{Magesan2011}
\bibinfo{author}{Magesan, E.}, \bibinfo{author}{Gambetta, J.~M.} \&
  \bibinfo{author}{Emerson, J.}
\newblock \bibinfo{journal}{\bibinfo{title}{Scalable and robust randomized
  benchmarking of quantum processes}}.
\newblock {\emph{\JournalTitle{Physical Review Letters}}}
  \textbf{\bibinfo{volume}{106}}, \bibinfo{pages}{180504},
  \doiprefix\url{10.1103/PhysRevLett.106.180504} (\bibinfo{year}{2011}).

\bibitem{cross2019}
\bibinfo{author}{Cross, A.~W.}, \bibinfo{author}{Bishop, L.~S.},
  \bibinfo{author}{Sheldon, S.}, \bibinfo{author}{Nation, P.~D.} \&
  \bibinfo{author}{Gambetta, J.~M.}
\newblock \bibinfo{journal}{\bibinfo{title}{Validating quantum computers using
  randomized model circuits}}.
\newblock {\emph{\JournalTitle{Physical Review A}}}
  \textbf{\bibinfo{volume}{100}}, \bibinfo{pages}{032328},
  \doiprefix\url{10.1103/PhysRevA.100.032328} (\bibinfo{year}{2019}).

\bibitem{Arute_2019_Sycamore}
\bibinfo{author}{Arute, F.} \emph{et~al.}
\newblock \bibinfo{journal}{\bibinfo{title}{Quantum supremacy using a
  programmable superconducting processor}}.
\newblock {\emph{\JournalTitle{Nature}}} \textbf{\bibinfo{volume}{574}},
  \bibinfo{pages}{505--510}, \doiprefix\url{10.1038/s41586-019-1666-5}
  (\bibinfo{year}{2019}).

\bibitem{erhard2019characterizing}
\bibinfo{author}{Erhard, A.} \emph{et~al.}
\newblock \bibinfo{journal}{\bibinfo{title}{Characterizing large-scale quantum
  computers via cycle benchmarking}}.
\newblock {\emph{\JournalTitle{Nature Communications}}}
  \textbf{\bibinfo{volume}{10}}, \bibinfo{pages}{5347},
  \doiprefix\url{10.1038/s41467-019-13068-7} (\bibinfo{year}{2019}).

\bibitem{proctor2022scalable}
\bibinfo{author}{Proctor, T.}, \bibinfo{author}{Rudinger, K.},
  \bibinfo{author}{Young, K.}, \bibinfo{author}{Sarovar, M.} \&
  \bibinfo{author}{Blume-Kohout, R.}
\newblock \bibinfo{journal}{\bibinfo{title}{Scalable randomized benchmarking of
  quantum computers using mirror circuits}}.
\newblock {\emph{\JournalTitle{Physical Review Letters}}}
  \textbf{\bibinfo{volume}{129}}, \bibinfo{pages}{150502},
  \doiprefix\url{10.1103/PhysRevLett.129.150502} (\bibinfo{year}{2022}).

\bibitem{Howard2014}
\bibinfo{author}{Howard, M.}, \bibinfo{author}{Wallman, J.},
  \bibinfo{author}{Veitch, V.} \& \bibinfo{author}{Emerson, J.}
\newblock \bibinfo{journal}{\bibinfo{title}{Contextuality supplies the `magic'
  for quantum computation}}.
\newblock {\emph{\JournalTitle{Nature}}} \textbf{\bibinfo{volume}{510}},
  \bibinfo{pages}{351--355}, \doiprefix\url{10.1038/nature13460}
  (\bibinfo{year}{2014}).

\bibitem{AQT_Marmot}
\bibinfo{title}{{Alpine Quantum Technologies} webpage about marmot.}
\newblock \bibinfo{howpublished}{\url{https://www.aqt.eu/products/marmot/}}.
\newblock \bibinfo{note}{Accessed: December 2025}.

\bibitem{Garnet_benchmark}
\bibinfo{author}{Abdurakhimov, L.} \emph{et~al.}
\newblock \bibinfo{title}{Technology and performance benchmarks of iqm's
  20-qubit quantum computer} (\bibinfo{year}{2024}).

\bibitem{CHSH_inequalities}
\bibinfo{author}{Clauser, J.~F.}, \bibinfo{author}{Horne, M.~A.},
  \bibinfo{author}{Shimony, A.} \& \bibinfo{author}{Holt, R.~A.}
\newblock \bibinfo{journal}{\bibinfo{title}{Proposed experiment to test local
  hidden-variable theories}}.
\newblock {\emph{\JournalTitle{Phys. Rev. Lett.}}}
  \textbf{\bibinfo{volume}{23}}, \bibinfo{pages}{880--884},
  \doiprefix\url{10.1103/PhysRevLett.23.880} (\bibinfo{year}{1969}).

\bibitem{Alsina_2016}
\bibinfo{author}{Alsina, D.}, \bibinfo{author}{Cervera, A.},
  \bibinfo{author}{Goyeneche, D.}, \bibinfo{author}{Latorre, J.~I.} \&
  \bibinfo{author}{Życzkowski, K.}
\newblock \bibinfo{journal}{\bibinfo{title}{Operational approach to {B}ell
  inequalities: Application to qutrits}}.
\newblock {\emph{\JournalTitle{Phys. Rev. A}}} \textbf{\bibinfo{volume}{94}},
  \bibinfo{pages}{032102}, \doiprefix\url{10.1103/PhysRevA.94.032102}
  (\bibinfo{year}{2016}).

\bibitem{Werner_2001}
\bibinfo{author}{Werner, R.~F.} \& \bibinfo{author}{Wolf, M.~M.}
\newblock \bibinfo{journal}{\bibinfo{title}{All-multipartite {B}ell-correlation
  inequalities for two dichotomic observables per site}}.
\newblock {\emph{\JournalTitle{Phys. Rev. A}}} \textbf{\bibinfo{volume}{64}},
  \bibinfo{pages}{032112}, \doiprefix\url{10.1103/PhysRevA.64.032112}
  (\bibinfo{year}{2001}).

\bibitem{Aspect1982}
\bibinfo{author}{Aspect, A.}, \bibinfo{author}{Grangier, P.} \&
  \bibinfo{author}{Roger, G.}
\newblock \bibinfo{journal}{\bibinfo{title}{Experimental realization of
  {E}instein-{P}odolsky-{R}osen-{B}ohm {G}edankenexperiment: A new violation of
  {B}ell's inequalities}}.
\newblock {\emph{\JournalTitle{Phys. Rev. Lett.}}}
  \textbf{\bibinfo{volume}{49}}, \bibinfo{pages}{91--94},
  \doiprefix\url{10.1103/PhysRevLett.49.91} (\bibinfo{year}{1982}).

\bibitem{Hensen_2015}
\bibinfo{author}{Hensen, B.} \emph{et~al.}
\newblock \bibinfo{journal}{\bibinfo{title}{Loophole-free {B}ell inequality
  violation using electron spins separated by 1.3 kilometres}}.
\newblock {\emph{\JournalTitle{Nature}}} \textbf{\bibinfo{volume}{526}},
  \bibinfo{pages}{682--686}, \doiprefix\url{10.1038/nature15759}
  (\bibinfo{year}{2015}).

\bibitem{Significant-Loophole-Free-2015}
\bibinfo{author}{Giustina, M.} \emph{et~al.}
\newblock \bibinfo{journal}{\bibinfo{title}{Significant-loophole-free test of
  {B}ell's theorem with entangled photons}}.
\newblock {\emph{\JournalTitle{Phys. Rev. Lett.}}}
  \textbf{\bibinfo{volume}{115}}, \bibinfo{pages}{250401},
  \doiprefix\url{10.1103/PhysRevLett.115.250401} (\bibinfo{year}{2015}).

\bibitem{Strong-Loophole-Free-2015}
\bibinfo{author}{Shalm, L.~K.} \emph{et~al.}
\newblock \bibinfo{journal}{\bibinfo{title}{Strong loophole-free test of local
  realism}}.
\newblock {\emph{\JournalTitle{Phys. Rev. Lett.}}}
  \textbf{\bibinfo{volume}{115}}, \bibinfo{pages}{250402},
  \doiprefix\url{10.1103/PhysRevLett.115.250402} (\bibinfo{year}{2015}).

\bibitem{Rosenfeld2017}
\bibinfo{author}{Rosenfeld, W.} \emph{et~al.}
\newblock \bibinfo{journal}{\bibinfo{title}{Event-ready {B}ell test using
  entangled atoms simultaneously closing detection and locality loopholes}}.
\newblock {\emph{\JournalTitle{Phys. Rev. Lett.}}}
  \textbf{\bibinfo{volume}{119}}, \bibinfo{pages}{010402},
  \doiprefix\url{10.1103/PhysRevLett.119.010402} (\bibinfo{year}{2017}).

\bibitem{Schmied2016}
\bibinfo{author}{Schmied, R.} \emph{et~al.}
\newblock \bibinfo{journal}{\bibinfo{title}{{B}ell correlations in a
  {B}ose-{E}instein condensate}}.
\newblock {\emph{\JournalTitle{Science}}} \textbf{\bibinfo{volume}{352}},
  \bibinfo{pages}{441--444}, \doiprefix\url{10.1126/science.aad8665}
  (\bibinfo{year}{2016}).

\bibitem{Baccari_2019}
\bibinfo{author}{Baccari, F.} \emph{et~al.}
\newblock \bibinfo{journal}{\bibinfo{title}{Bell correlation depth in many-body
  systems}}.
\newblock {\emph{\JournalTitle{Phys. Rev. A}}} \textbf{\bibinfo{volume}{100}},
  \bibinfo{pages}{022121}, \doiprefix\url{10.1103/PhysRevA.100.022121}
  (\bibinfo{year}{2019}).

\bibitem{grover1996fast}
\bibinfo{author}{Grover, L.~K.}
\newblock \bibinfo{title}{A fast quantum mechanical algorithm for database
  search}.
\newblock In \emph{\bibinfo{booktitle}{Proceedings of the twenty-eighth annual
  ACM symposium on Theory of computing}}, \bibinfo{pages}{212--219},
  \doiprefix\url{10.1145/237814.237866} (\bibinfo{year}{1996}).

\bibitem{kitaev1995quantum}
\bibinfo{author}{Kitaev, A.~Y.}
\newblock \bibinfo{journal}{\bibinfo{title}{Quantum measurements and the
  abelian stabilizer problem}}.
\newblock {\emph{\JournalTitle{arXiv preprint quant-ph/9511026}}}
  (\bibinfo{year}{1995}).
\newblock \eprint{9511026}.

\bibitem{bernstein1997quantum}
\bibinfo{author}{Bernstein, E.} \& \bibinfo{author}{Vazirani, U.}
\newblock \bibinfo{journal}{\bibinfo{title}{Quantum complexity theory}}.
\newblock {\emph{\JournalTitle{SIAM Journal on Computing}}}
  \textbf{\bibinfo{volume}{26}}, \bibinfo{pages}{1411--1473},
  \doiprefix\url{10.1137/S0097539796300921} (\bibinfo{year}{1997}).

\bibitem{gretton2012kernel}
\bibinfo{author}{Gretton, A.}, \bibinfo{author}{Borgwardt, K.~M.},
  \bibinfo{author}{Rasch, M.~J.}, \bibinfo{author}{Sch{\"o}lkopf, B.} \&
  \bibinfo{author}{Smola, A.}
\newblock \bibinfo{journal}{\bibinfo{title}{A kernel two-sample test}}.
\newblock {\emph{\JournalTitle{The journal of machine learning research}}}
  \textbf{\bibinfo{volume}{13}}, \bibinfo{pages}{723--773}
  (\bibinfo{year}{2012}).

\bibitem{psb25}
\bibinfo{author}{{Piskor}, T.} \emph{et~al.}
\newblock \bibinfo{journal}{\bibinfo{title}{{Simulation and Benchmarking of
  Real Quantum Hardware}}}.
\newblock {\emph{\JournalTitle{arXiv e-prints}}}
  \bibinfo{pages}{arXiv:2508.04483}, \doiprefix\url{10.48550/arXiv.2508.04483}
  (\bibinfo{year}{2025}).

\bibitem{IQM_Garnet}
\bibinfo{author}{{IQM Academy}}.
\newblock \bibinfo{title}{Garnet qpu benchmarking page}.
\newblock
  \bibinfo{howpublished}{\url{https://www.iqmacademy.com/qpu/benchmarking/garnet/}}
  (\bibinfo{year}{2025}).
\newblock \bibinfo{note}{Accessed: December 2025}.

\bibitem{Google_XEB}
\bibinfo{title}{{Google} use of xeb as a calibration tool}.
\newblock
  \bibinfo{howpublished}{\url{https://quantumai.google/cirq/noise/qcvv/xeb_theory}}.
\newblock \bibinfo{note}{Accessed: December 2025}.

\end{thebibliography}

\end{document}